\documentclass[a4paper,fleqn,usenatbib]{mnras}

\usepackage[T1]{fontenc}
\usepackage{ae,aecompl}
\usepackage{empheq}

\usepackage{graphicx}	% Including figure files
\usepackage{amsmath}	% Advanced maths commands
\usepackage{amssymb}	% Extra maths symbols

\usepackage{amsfonts}
\usepackage{mathrsfs} % $\mathscr{L}$   
\usepackage{subcaption}
\title{Concentration, Ellipsoidal Collapse, and the Densest Dark Matter Haloes}

\author[C. Okoli and N. Afshordi]{Chiamaka Okoli$^{1,2}$\thanks{E-mail:c2okoli@uwaterloo.ca} and Niayesh Afshordi$^{1,2}$\thanks{E-mail:nafshordi@pitp.ca}\\
$^{1}$Perimeter Institute for Theoretical Physics, 31 Caroline Street North, Waterloo, ON N2L 2Y5, Canada\\
$^{2}$Department of Physics and Astronomy, University of Waterloo, 200 University Avenue West, Waterloo,ON  N2L 3G1, Canada}

\date{Accepted XXX. Received YYY; in original form ZZZ}

\pubyear{2015}

\begin{document}

\label{firstpage}
\pagerange{\pageref{firstpage}--\pageref{lastpage}}
\maketitle

\begin{abstract}
The smallest dark matter haloes are the first objects to form in the hierarchical structure formation of cold dark matter (CDM) cosmology and are expected to be the densest and most fundamental building blocks of CDM structures in our universe. Nevertheless, the physical characteristics of these haloes have stayed illusive, as they remain well beyond the current resolution of N-body simulations (at redshift zero). However, they dominate the predictions (and uncertainty) in expected dark matter annihilation signal, amongst other astrophysical observables. Using the conservation of total energy and the ellipsoidal collapse framework, we can analytically find the mean and scatter of concentration $c$  and 1-D velocity dispersion $\sigma_{\rm 1d}$ for haloes of different virial mass $M_{200}$. Both $c$ and $\sigma_{\rm 1d}/M_{200}^{1/3}$ are in good agreement with numerical results within the regime probed by simulations -- slowly decreasing functions of mass that approach constant values at large masses. In particular, the predictions for the 1-D velocity dispersion of cluster mass haloes are surprisingly robust as the inverse heat capacity of cosmological haloes crosses zero at $M_{200} \sim 10^{14} M_\odot$. However, we find that current extrapolations from simulations to smallest CDM haloes dramatically depend on the assumed profile (e.g. NFW vs. Einasto) and fitting function, which is why theoretical considerations, such as the one presented here, can significantly constrain the range of feasible predictions. 

\end{abstract}

\section{Introduction}\label{sec:intro}

In the study of structure formation in the universe, dark matter haloes form the bedrock from which galaxies and clusters of galaxies grow \citep{b1}. This is evident in the cold dark matter (CDM) universe where haloes build up hierarchically from the collapse of primordial density perturbations out of an expanding background. Dark matter haloes have been extensively studied  in the literature using both N-body simulations and  analytical approaches \citep[][and references therein]{b8,b2,b3,b6,b7,b4,b5}. Over the years, N-body simulations and analytical models have improved our understanding of the properties and structure of dark matter haloes. Thus, a range of density profiles have been found for CDM haloes which include the widely studied NFW profile \citep{b2},
\begin{equation}
\rho(r) = \frac{\rho_s}{r/r_s\left(1+r/r_s\right)^2},
\end{equation}
and the Einasto profile \citep{c36,c34}
\begin{equation}
\rho(r) = \rho_s \exp\left\{ -\frac{2}{\alpha}\left[\left(\frac{r}{r_s}\right)^{\alpha} - 1\right]\right\}
\end{equation}
where $\rho_s$  and $r_s$ are the scale parameters and $\alpha$ is a shape parameter for the Einasto profile. Note that:
\begin{equation}
\frac{d \ln\rho(r)}{d \ln r} = -2 {\rm ~~~for~~~}r=r_s,
\end{equation}
for both parametric forms, which is why $r_{-2}$ is also used instead of $r_s$ in the literature.

An alternative parametrization of the density profile involves the halo mass and concentration. The halo concentration characterizes the halo central density, and may be defined by 
\begin{equation}
c_{200} \equiv \frac{r_{200}}{r_s}, 
\end{equation}
where $r_{200}$ is the radius of the sphere within which the mean halo density is 200 times the {\it critical} density of the universe. N-body simulations have successively shown that these two parameters are correlated with each other -- the halo concentration is a decreasing function of mass, with a redshift dependence (decrease with increasing redshift) at fixed masses \citep{b2,b3,b39,b12,b38,b29,b24,b13,c2}. This relation reflects the different formation times of haloes \citep{b34}, with low mass haloes forming earlier. Previous studies show that the dark matter halo concentrations are lognormally distributed at fixed halo masses \citep{b12,b24}. 

Current knowledge of the concentration--mass relation has been largely based on N-body simulations. N-body simulations are quite expensive and time consuming. In addition, at fixed mass, the concentration--mass relation of a halo depends on the values of the cosmological parameters -- the rms amplitude of linear matter fluctuation, the matter density, the spectral index, the Hubble's constant, and the baryon density \citep{b33,b15}. For this reason, mass concentration relations from N-body simulations are limited to the values of the cosmological parameters used, as well as masses resolved, in the simulation. To circumvent these challenges, the need for an analytical framework to link the mass and concentration of dark matter haloes cannot be overemphasized. Several fitting formulae, mostly calibrated by simulations, exist in the literature. For example, \citet{b29} proposed a $c(m)$ model dependent on the variance of linear density fluctuations in a sphere of mass $m$, $\sigma^2(m)$. Their results, which agree with the results of their simulations, however,  show a surprising and controversial upturn in the $c(m)$ relation at large masses. \citet{b13} recently proposed a concentration--mass-redshift relation of haloes using the mass accretion history of haloes. In particular, they used both simulated and analytic estimates of the mass accretion histories to determine the concentration of haloes. A comparison of the results of a number of these models to ours is examined at the end of this paper.

The concentration--mass relation gives deep insight into the formation and structure of haloes. Accurate concentration--mass relations can also be used to search for the elusive dark matter particles by placing limits on the dark matter annihilation flux from (sub)structures e.g \citep{b16}. Indeed, the latter is dominated by smallest haloes not resolved in N-body simulations. The main objective of this work is to derive an analytical concentration--mass relation for haloes, using physical principles of {\it energy conservation } and {\it ellipsoidal collapse}, which gives theoretical insight {\it and} clarity into the origin of such a relation.  In what follows, we derive an analytical relation between the concentration and the mass of haloes using the conservation of the total energy of a comoving region \citep{b23} through the gravitational collapse with a modified relation, accounting for the ellipsoidal collapse. Our results are then compared with results obtained from simulations to validate the veracity of our assumptions. Most significantly however, as our framework is rooted in physical principles, it can be applied and trusted well beyond the regime probed by simulations. 

As another useful application of this framework, we shall find a robust prediction for velocity dispersion of dark matter haloes. The small scatter, as well as cosmology independence of the velocity dispersion-mass relationship for cluster mass haloes \citep{c5} can then be understood through a simple and elegant energetic argument.

The outline of this work is as follows: In the next section, we define our random variables, their probability distribution and the initial total energy of a spherical region with random Gaussian initial conditions. The mean and dispersion of our random variables are then related to the concentration through our assumption of energy conservation. Section \ref{sec:HCM} incorporates the possibility of non-sphericity using the ellipsoidal collapse model of \citet{b35}. We then compare the predictions of our model to results from simulations and their naive extrapolations. Section \ref{sec:sigma} briefly discusses our predictions for halo velocity dispersions. Finally, we discuss our results and summarize our findings in Section \ref{conclude}. The impatient reader can jump to the last section to find practical and concise fitting functions to our results.  

Throughout this paper, we assume $\Lambda$CDM cosmology with Gaussian initial conditions. 

\section{Halo statistics using the spherical collapse model}\label{sec:sphere}
The spherical collapse model \citep{b26} describes the formation of structure from the collapse of a spherical region perturbed in density. In general, non-linear gravitational dynamics is difficult to deal with analytically; however, the assumption of the symmetry of the system simplifies the dynamics. For simplicity and since haloes are usually approximated as spherical systems, we consider a spherical overdensity field. We also ignore the tidal effects of neighbouring density perturbations upon the evolution of the isolated homogeneous spherical density perturbation. In what follows, we calculate the initial energy of this region and its virialization time. These are related to random variables whose probability distribution will play a key role in predicting the mean and dispersion of halo concentrations.
\subsection{Initial Energy}
\label{subsec:energy}
The total energy of an isolated system at a given time can be given as the sum of the kinetic energy and potential energy at that time. Thus, we derive the initial kinetic energy and potential energy of a spherical volume prior to collapse. For the kinetic energy of the region, we note that the velocity can be written as a function of gravitational potential $\phi_i$ \footnote{$i$ here stands for the initial time} \citep{b27},
\begin{equation}
\textbf{v} = H_i\textbf{x} - \frac{2}{3H_i}\nabla\phi_i,
\label{first}
\end{equation}
where $H_i$ is the Hubble constant at the initial time. However, note that our results are independent of this choice of initial time, as long as it is in the linear regime. Writing the initial density of the region as a perturbation to the initial mean density of the universe  $$\rho({\bf x},t_i)  = \rho_i[1+\delta_i({\bf x})],$$  the kinetic energy, to linear order in the perturbation, is then given by
\begin{equation}
K_i = \frac{1}{2}\rho_i\int{(H_i^2|\textbf{x}|^2 - \frac{4}{3}\textbf{x}\cdot\nabla\phi_i + H_i^2|\textbf{x}|
^2\delta_i)d^3x}.
\end{equation}
The perturbation $\delta_i$ can be substituted in favour of the gravitational potential with the aid of the Poisson equation, then simplified further using the Friedmann equation to give
\begin{eqnarray}
K_i = \frac{1}{2}\rho_i\int{(H_i^2|\textbf{x}|^2 + \frac{2}{3}x^2\nabla^2\phi_i-  
\frac{4 }{3}\textbf{x}\cdot\nabla\phi_i)d^3x}\\
 = \frac{1}{2}\rho_i\int{(H_i^2|\textbf{x}|^2 + \frac{4}{3}x^2\nabla^2\phi_i)d^3x} - \frac{1}{3}\rho_i\oint{x^2\nabla\phi_i\cdot d\textbf{a}},
\label{eq:div}
\end{eqnarray}
where we have used the divergence theorem in simplifying Equation \ref{eq:div}. Neglecting deviations from spherical symmetry at the boundary, we have
$$\nabla\phi_i = \frac{G \delta M}{R_i^2}\hat{r} = \hat{r} \frac{G \rho_i}{R_i^2}\int{\delta_id^3x}.$$ Finally, the kinetic energy is given as
\begin{equation}
K_i = \frac{4\pi G \rho_i^2}{3}\int_0^{R_i}{\big[x^2 + \delta_i(2x^2 - R_i^2) \big]d^3x}.
\end{equation}
Clearly, the kinetic energy is the sum of that expected from spherical volume with mean density and that due to the perturbation. 
Similarly, the initial gravitational potential energy of the spherical region can be expressed as
\begin{equation}
U_i = -\frac{G \rho_i^2}{2}\int\int{\frac{\left[1+ \delta_i({\bf x}_1)\right]\left[1+ \delta_i({\bf x}_2)\right]}{|{\bf x}_1 - {\bf x}_2|}}d^3x_1d^3x_2
\end{equation}
To linear order in $\delta_i$ and using the symmetry under the interchange of $x_1$ and $x_2$, the potential energy is re-expressed as 
\begin{eqnarray}
U_i &=& -\frac{G \rho_i^2}{2}\int{[1 + 2 \delta_i({\bf x}_1)]d^3x_1}\int{\frac{d^3x_2}{|{\bf x}_1 - {\bf x}_2|}} \nonumber\\
 &=& -\frac{4\pi G}{3}\rho_i^2\int_0^{R_i}{(1 + 2\delta_i)\left(\frac{3R_i - x^2}{4}\right)d^3x},
\label{eq:pot}
\end{eqnarray}
where we arrived at the last part of Equation \ref{eq:pot} by taking the second integral in a spherical volume. The initial energy, which is the sum of the initial kinetic energy and potential energy, is then given by
\begin{eqnarray}
E_i &=& -\frac{10}{3}\pi G \rho_i^2 R_i^5\left[\int_0^{R_i}{\delta_i({\bf x})\frac{1}{R_i^3}\left(1-\frac{x^2}{R_i^2}\right)d^3x}\right]\nonumber\\
    &=& -\frac{10}{3}\pi G \rho_i^2 R_i^5 B. 
\label{eq:energy}
\end{eqnarray}
The parameter $B$ is thus defined as 
\begin{equation}
B \equiv \int_0^{R_i}{\delta_i({\bf x})\frac{1}{R_i^3}\left(1-\frac{x^2}{R_i^2}\right)d^3x}.
\end{equation}
 It is important to note that the integral in the definition of $B$ is a three dimensional integral whose domain is within a sphere of radius $R_i$, and we assume $\delta_i({\bf x})$ is a random Gaussian field. Physically, $B$ is the linear overdensity in the inner regions of a spherical region of initial radius $R_i$. For a  given halo with initial radius $R_i$, the total initial energy of the halo fixes $B$.

\subsection{Virialization Time}
In the linear regime, density perturbations grow linearly with scale factor until they reach a critical value, after which they turn around from the uniform expansion of the universe and collapse to form virialized dark matter haloes. Various relaxation processes occur during the collapse of a spheroid from rest which prevents the object from collapsing to a point. However, one can safely assume that the collapsing object virializes at around half its radius at turnaround \citep{b26}. To examine this, consider a test particle with unit mass on the boundary of a spherical region of radius $R_i$, and initial mass, $M$, the total energy $e$ of the particle is given as $$e = \frac{\textbf{v}_i^2}{2}-\frac{G M}{R_i}.$$ Assuming that the collapse time of the particle is approximately the same time necessary for the particle to be virialized, the collapse time $t$ can be written as
\begin{equation}
t = \frac{2\pi G M}{(-2e)^{3/2}}.
\end{equation}
With the mass, $M$ interior to the test particle assumed to be virialized at $t$, we then relate the initial density to the collapse time through 
\begin{equation}
-2e = \frac{5}{4\pi}H_i^2R_i^2\int_0^{R_i}{\delta_i\frac{d^3x}{R_i^3}}.
\end{equation}
Using the Friedmann equation and that $M$ = $(4/3)\pi R_i^3\rho_i$, we obtain
\begin{equation}
A \equiv \int_0^{R_i}{\delta_i\frac{d^3x}{R_i^3}} = \frac{2}{5}\left(\frac{3\pi^4}{t^2 G \rho_i}\right)^{1/3}
\label{eq:A}
\end{equation} 
Therefore, $A$ relates the initial density perturbation of a spherical region to the virialization/formation time of a dark matter halo. Physically, in contrast to $B$, $A$ is the mean linear overdensity of a region before collapse. Notice that Eq. (\ref{eq:A}) is equivalent to the standard spherical collapse threshold, when translated into linear overdensity today: $A \rightarrow \frac{4\pi}{3}  \delta_{\rm sc} \simeq   \frac{4\pi}{3} \times 1.686$ for Einstein-de Sitter cosmology \citep{b26}. 

\subsection{Probability Distribution of the Parameters A and B}\label{sec:prob_ab}
The parameters $A$ and $B$ are dependent on the linear density field which is a random Gaussian field. In this subsection, we study the resulting joint probability distribution of $A$ and $B$. Recall that we had defined $$A \equiv \int_0^{R_i}{\delta_i({\bf x})\frac{d^3x}{R_i^3}},$$ and $$B \equiv \int_0^{R_i}{\delta_i({\bf x})\frac{1}{R_i^3}\left(1-\frac{x^2}{R_i^2}\right)d^3x}.$$  
Assuming Gaussian statistics for the linear density field, the probability distribution function for $A$ and $B$ takes the form:
\begin{eqnarray}
&&P(A,B)dA dB 
= \nonumber\\ &&\frac{1}{2\pi \sqrt{L}}\exp{\left[- \frac{1}{2L}\left(\left\langle B^2 \right\rangle A^2 + \left\langle A^2 \right\rangle B^2 + 2 \left\langle AB \right\rangle AB\right)\right]}dA dB, \nonumber\\ \label{eq:gaussian}
\end{eqnarray}
where $L = \left\langle A^2\right\rangle \left\langle B^2\right\rangle - \left\langle AB \right\rangle ^2$. To determine this distribution, the values of the various spectra, $\left\langle B^2\right\rangle$, $\left\langle A^2\right\rangle$, and $\left\langle AB \right\rangle$ will have to be evaluated. To this end, we rewrite $A$ as  
$$A= \int \frac{d^3x}{R_i^3}\delta_i({\bf x}) U({\bf x}),$$ where $U({\bf x}) \equiv \theta(R_i-|{\bf x}|)$, is a step function.
Evaluating the average of the square of $A$ in the Fourier space gives
\begin{equation}
\left\langle A^2\right\rangle = \frac{1}{2\pi^2}\int_0^{\infty}{\frac{k^2 dk}{R_i^6}\left|\widetilde{U}(k)\right|^2 P(k)},
\end{equation}
where $\widetilde{U}(k)$ is the Fourier transform of $U(x)$ and the power spectrum, $P(k)$ is defined by $$(2\pi)^3 \delta^3 ({\bf k}+{\bf k'}) P(k) = \left\langle \delta_{\bf k} \delta_{\bf k'} \right\rangle.$$ The power spectrum is the Fourier transform of the spatial correlation function, which is invariant under spatial translations and rotations. It is pertinent to note that this invariance is expected since the cosmological field is spatially homogeneous and isotropic. We should note that the standard definition of the variance of spherical top-hat linear density perturbations is related to $\left\langle A^2\right\rangle$ by
\begin{equation}
\sigma^2(M) = \left(4\pi \over 3\right)^{-2} \left\langle A^2\right\rangle,
\end{equation}
after extrapolating $A$ using linear growth to today. 

\begin{figure}
 \centering
\includegraphics[width=1.0\columnwidth]{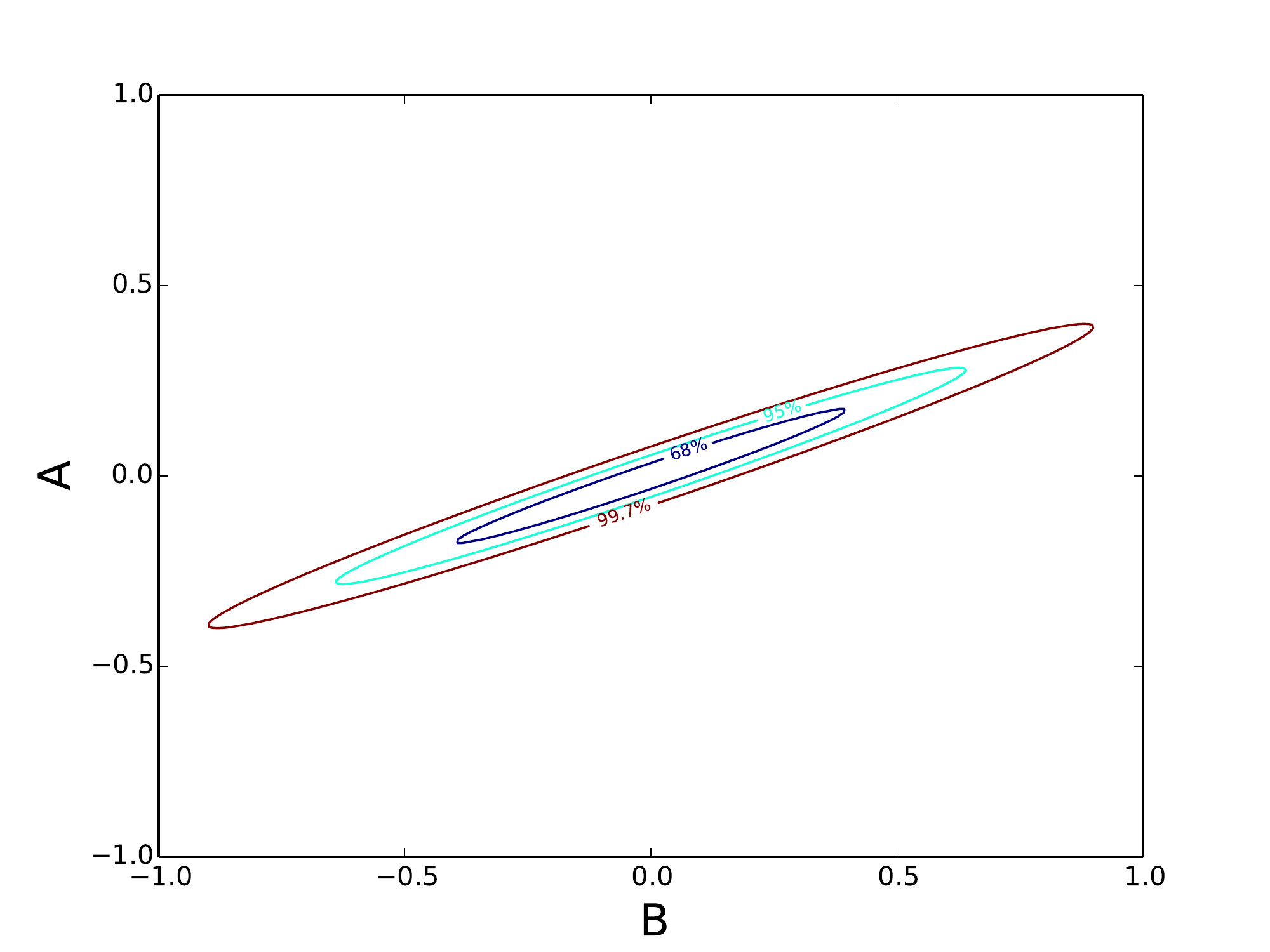}
\caption{A typical joint probability distribution of the parameters $A$ and $B$ (extrapolated to today using linear growth) for a spherical region of comoving radius, $R = 8h^{-1}$ Mpc ($M_{200} \simeq 3 \times 10^{14} M_\odot$). The contours show the  0.68, 0.95, 0.99 confidence regions in the distribution.}
\label{fig:probability}
\end{figure}

Now, rewriting $B$ as  
\begin{equation}
B = \int \frac{d^3x}{R_i^3}\delta_i({\bf x})W({\bf x}),
\end{equation}
 with $$W({\bf x}) \equiv \left(1 - \frac{|{\bf x}|^2}{R_i^2}\right) \theta(R_i-|{\bf x}|).$$ Similarly, the average of $B^2$ 
 \begin{equation}
\left\langle B^2\right\rangle = \frac{1}{2\pi^2}\int_0^{\infty}{\frac{k^2 dk}{R_i^6}\left|{\widetilde W}(k)\right|^2 P(k)},
\end{equation}
while $\left\langle AB \right\rangle$ is given by
\begin{equation}
\left\langle AB \right\rangle = \frac{1}{2\pi^2}\int_0^{\infty}{\frac{k^2 dk}{R_i^6} \widetilde{U}(k) \widetilde{W}(k) P(k)}.
\end{equation}
For Gaussian random fields, the power spectrum completely specifies all other correlations of the field. In the linear regime, Fourier modes evolve independently, thus a Gaussian fluctuation field remains Gaussian. The joint probability distribution of $A$ and $B$, which shows how the initial energy of a spherical region ($R = 8h^{-1} Mpc$) correlates with its collapse time is shown in Fig. \ref{fig:probability}.
\subsection{Jeans Equation}
The spherical Jeans equation relates the integrated mass, $M(r)$ of a spherically symmetric, dispersion-supported, collisionless system to its radial velocity dispersion, $\sigma(r)$ and mass density, $\rho(r)$, under the assumption of dynamical equilibrium. In a generalized coordinate system, the Jeans equation governing a system in dynamical equilibrium \citep{b28} is given as
\begin{equation}
\frac{\partial}{\partial t}\left( \rho \left \langle v_j \right\rangle \right) + \frac{\partial}{\partial x_i}\left(\rho \left \langle v_i v_j \right \rangle \right) + \frac{\partial \Phi}{\partial x_i}\rho \delta_{ij} = 0.
\label{eq:jean}
\end{equation}
Note that if the density, $\rho$, and the potential, $\Phi$, are known, then this is a system of three equations with six unknown second order velocity moments. To close the system of equations, we assume that the mean velocity (streaming motion) of the particles in any direction, and the velocity covariance among different components are zero. 

Evaluating Equation \ref{eq:jean} in spherical coordinates, the spherical Jeans equation in the static limit is given by
\begin{equation}
\frac{\partial \left(\rho \left\langle v^2_r \right\rangle\right)}{\partial r} + \frac{2}{r} \rho \left\langle v_r^2 \right\rangle - \frac{\rho}{r}\left\langle v_{\theta}^2 \right\rangle - \frac{\rho}{r}\left\langle v_{\phi}^2 \right\rangle = - \rho \frac{d \Phi}{dr}.
\end{equation}
In terms of the dispersion, $\sigma_i^2$, the velocity anisotropy parameter is defined as, 
\begin{equation}
\beta = 1 - \left(\frac{\sigma_t}{\sigma_r}\right)^2,
\label{eq:beta}
\end{equation}
where $\sigma^2_t \equiv \frac{\sigma_{\theta}^2 + \sigma_{\phi}^2}{2}$ and $\sigma_r$ are the tangential and radial component of the velocity dispersion respectively. Thus, the familiar Jeans equation for a spherically symmetric system in equilibrium is given by,  
\begin{equation}
\frac{d}{dr}\left(\rho \sigma_r^2 \right) + \frac{2}{r}\rho \beta\sigma_r^2 = - \rho \frac{d\Phi}{dr}.
\label{eq:jeans}
\end{equation}
The velocity anisotropy parameter $\beta$  measures the deviation of the motion of a system of particles from isotropy. For purely circular orbits, $\sigma_r = 0$, $\beta = - \infty$, whereas for purely radial orbits $\sigma_{\theta} = \sigma_{\phi} = 0$, $\beta = 1$. For isotropic motion, $\sigma_r$ = $\sigma_{\theta}$ = $\sigma_{\phi} $, $\beta = 0$. 

\begin{figure}
\centering
\includegraphics[width=1.0\columnwidth]{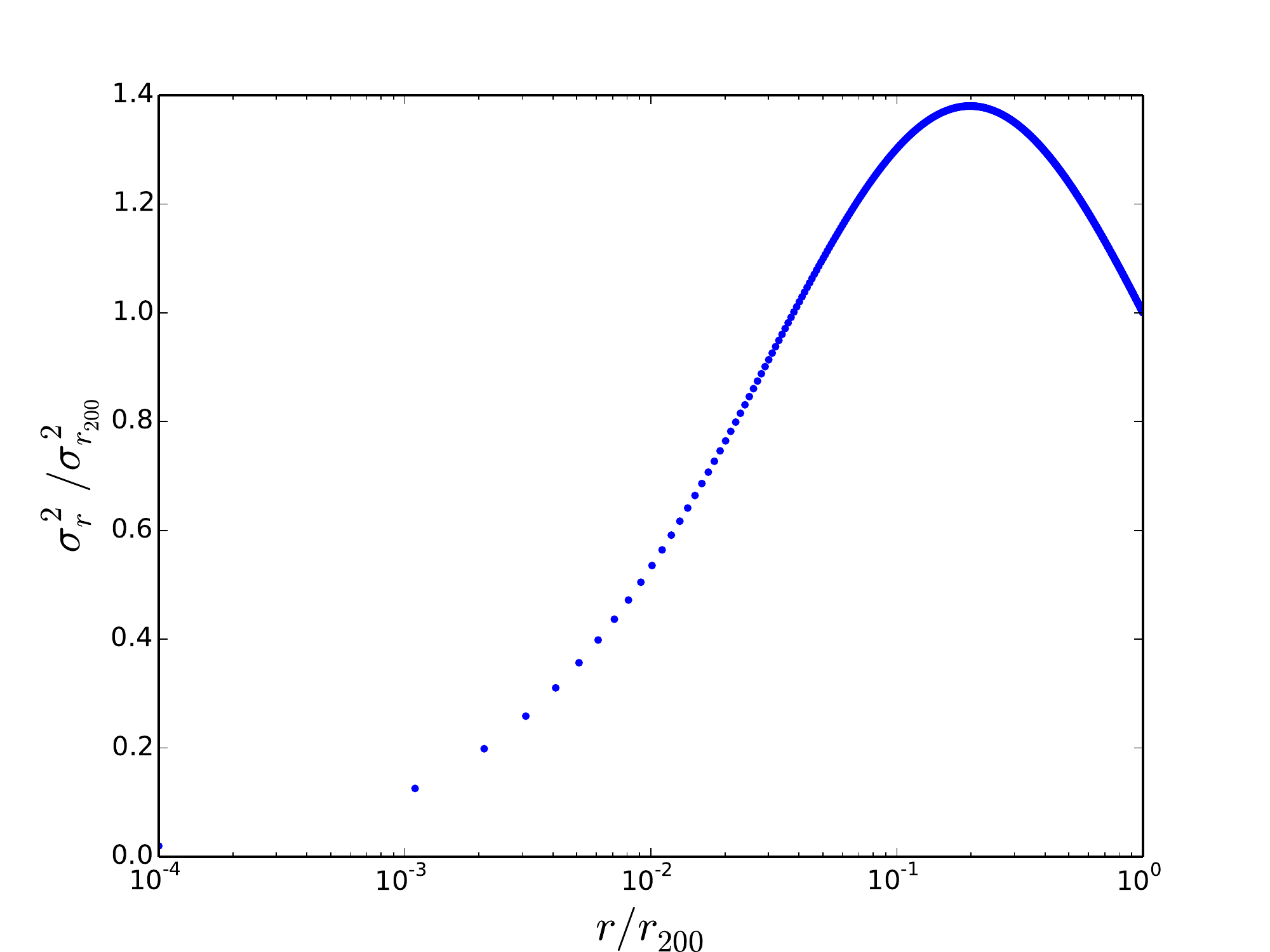}
\caption{Radial dependence of the radial velocity dispersion, based on solving the Jeans equation with anisotropy parameter (\ref{eq:hansen}), for an NFW halo with $c=4$. The radial distance is in units of virial radius, $r_{200}$ while the radial velocity dispersion is normalized to its value at the virial radius.}
\label{fig:vel}
\end{figure}

The Jeans equation can be solved for the dependence of the radial velocity dispersion on radius for a fixed density profile and velocity anisotropy profile. While it is easy to solve Jeans equation for isotropic velocity dispersions, simulations show that haloes are not isothermal and thus have radially dependent velocity anisotropy profile. Various studies have revealed that this velocity anisotropy profile is a nonzero radially varying function, with a value close to $0$ in the centre to approximately $ 0.4$ in the outer regions of the halo \citep{b7,b17,b30,b20,b31,b18,b19}. \citet{b20} likened this relation to the ratio of the gravitational potential energy to the kinetic energy within the NFW scale radius for haloes with the NFW-like density profiles. For the purpose of our calculations, we assume that the anisotropic velocity dispersion parameter $\beta(r)$ is linearly related to the logarithmic slope of the density profile, $\frac{d\ln{\rho(r)}}{d\ln{r}}$ \citep{b17,b20} in an almost universal way by 
\begin{equation}
\beta(r) = 1 - 1.15 \left[1+\frac{1}{6}\frac{d\ln{\rho(r)}}{d\ln{r}} \right].
\label{eq:hansen}
\end{equation}
The solution to the Jeans equation is thus given by 
\begin{eqnarray}
\rho(r) \sigma_r^2(r) = \frac{\int_r^{\infty}dr' F(r') \rho(r') \frac{d \phi(r')}{d r'}}{F(r)}, \\
F(r) \equiv \exp\left(\int_0^r2\frac{\beta(r')}{r'}\right),
\end{eqnarray}
where we have assumed $\rho(r) \sigma_r^2(r) \rightarrow 0$ as $r\rightarrow \infty$.
Numerically calculating the velocity dispersion as a function of radial distance for a halo with an $NFW$ profile, a concentration of 4, and an anisotropy profile given by Equation \ref{eq:hansen} yields a velocity dispersion profile shown in Fig. \ref{fig:vel}. The radial distance is in units of virial radius, $r_{200}$, while the radial velocity dispersion is normalized to its value at the virial radius. The velocity dispersion increases rapidly with radius at small radii (near the minimum of the potential), reaches a peak and then decreases outwards.
%An important caveat to the results obtained is the assumption that the density profile and velocity anisotropy profile of halos have the same functional form beyond the virial radius up to infinity.

Although haloes are approximated to be in equilibrium at the virial radius, there is a continuous infall of matter onto the halo boundary and therefore a considerable amount of surface pressure at the boundary \citep{b32}. Integrating the Jeans equation (\ref{eq:jeans}) over a spherical region, the expected correction to the virial theorem due to infalling matter at the boundary  is then given as
\begin{equation}
2K+U \simeq 4 \pi r^3 \rho \sigma^2_r \bigg|_{ r=r_{200}} .
\label{eq:mu}
\end{equation}
The first term on the LHS is twice the total kinetic energy of the region, while the second term is the gravitational potential energy. The RHS in the Eq. (\ref{eq:mu}) appears due to the  non-vanishing external pressure at the boundary. For a vanishing pressure on the boundary, we have the familiar virial relation -- the sum of the potential energy and twice the kinetic energy is zero. 
%We can then assume that the  boundary term (related to the external surface pressure) is proportional to the gravitational  potential energy, $U$ by $$3P_{\rm ext}V = \omega U.$$ 
We can now solve the Jeans equation (\ref{eq:jeans}), using the anisotropy parameter $\beta(r)$  (\ref{eq:hansen}) and virial theorem (\ref{eq:mu}), for any density profile $\rho(r)$ (e.g. NFW or Einasto) to find: 
\begin{equation}
\omega \equiv -\frac{4\pi r^3\rho \sigma_r^2}{U}\bigg|_{r=r_{200}}, \label{eq:omega}
\end{equation}
which only depends on the dimensionless parameters of the profile (e.g. concentration, and possibly $\alpha$ for Einasto profile). 
The parameter $\omega$ plays an important role in our prescription for the derivation of the concentration--mass relation of dark matter haloes, from conservation of energy.
\section{Halo Concentration and Mass}\label{sec:HCM}
\subsection{Spherical Collapse Model}
\label{subsec:sphere}
In this section, we derive an analytical relation between the concentration and mass of dark matter haloes by imposing that the total energy of a spherical region before collapse is equal to the total energy of the virialized halo formed from the collapse. This assumption is justified for spherical regions which are not coupled to the expansion of the background. 

Let us first define a dimensionless measure of the total energy of the halo, $E$, as:
\begin{equation}
y \equiv -\frac{4E}{3M_{200}}\left(\frac{1}{2\pi G M_{200} H}\right)^{2/3}.
\label{eq:yn}
\end{equation}

We can use the modified virial theorem (Equations \ref{eq:mu}-\ref{eq:omega}) to find the final energy $E_f$ of the virialized halo in terms of its density profile:  
\begin{eqnarray}
y_f &\equiv& -\frac{4E_f}{3M_{200}}\left(\frac{1}{2\pi G M_{200} H}\right)^{2/3} = -\frac{1}{3}\left(\frac{200}{\pi^2}\right)^{1/3}\frac{r_{200}(1-\omega)U}{GM_{200}^2} \nonumber \\ &=& \frac{1}{3}\left(\frac{200}{\pi^2}\right)^{1/3} (1-\omega) \int_0^1 \frac{m(<x)}{x} dm,
\label{eq:yc}
\end{eqnarray}
where $x$ and $m(<x)$ are the radius and enclosed mass in units of $r_{200}$ and $M_{200}$. 

After solving the Jeans equation to find $\omega$, as described in the last section, Equation \ref{eq:yc} gives $y_f$ in terms of concentration of the halo for any assumed halo profile. A fitting relation for $c(y_f)$ accurate to $10\%$ for $0.5<y_f<20$ and 0.1 $<$ $\alpha$ $<$ 0.52 for the Einasto profile is given by:
\begin{equation}
\log{c_{\rm Einasto}} \simeq a_1(\alpha) \ln{y_f}^2 + a_2(\alpha) \ln{y_f} + a_3(\alpha),
\label{eq:c_einasto}
\end{equation} 
where 
\begin{eqnarray}
a_1(\alpha) &=& -1.14 \alpha ^2 +0.89 \alpha  - 0.17\\ \nonumber 
a_2(\alpha) &=& 0.35 + 0.04 \alpha^{-1.13} \\ \nonumber
a_3(\alpha) &=& 0.50 \alpha^{-0.43} .
\label{eq:coeff}
\end{eqnarray}
 A similar relation accurate to $10\%$ also for $0.5<y_f<10$ using the NFW profile is 
\begin{equation}
\log{c_{\rm NFW}} \simeq 0.78\ln{y_f} + 1.09.
\label{eq:c_nfw}
\end{equation}

We have also already derived the initial energy, $ E_i$,  of the spherical region in Section \ref{subsec:energy}. Combining Equations (\ref{eq:energy}) and (\ref{eq:A}), we find:
\begin{equation}
y_i \equiv  -\frac{4E_i}{3M_{200}}\left(\frac{1}{2\pi G M_{200} H}\right)^{2/3} = \frac{B}{A}\left(Ht\right)^{-2/3},
\end{equation}
which only depends on cosmology and the statistics of the linear initial density field (through $A$ and $B$).

%
%Recall that $A$ and $B$ are related to the initial density perturbation and radius of the spherical region. The distribution of the initial density is approximated with multiple peaks within the radius, with the assumption of gaussian statistics. To evaluate equation $y_i$, we average over all the peaks within a radius (thus a fixed mass). 

\begin{figure}
\centering
\includegraphics[width=1.0\columnwidth]{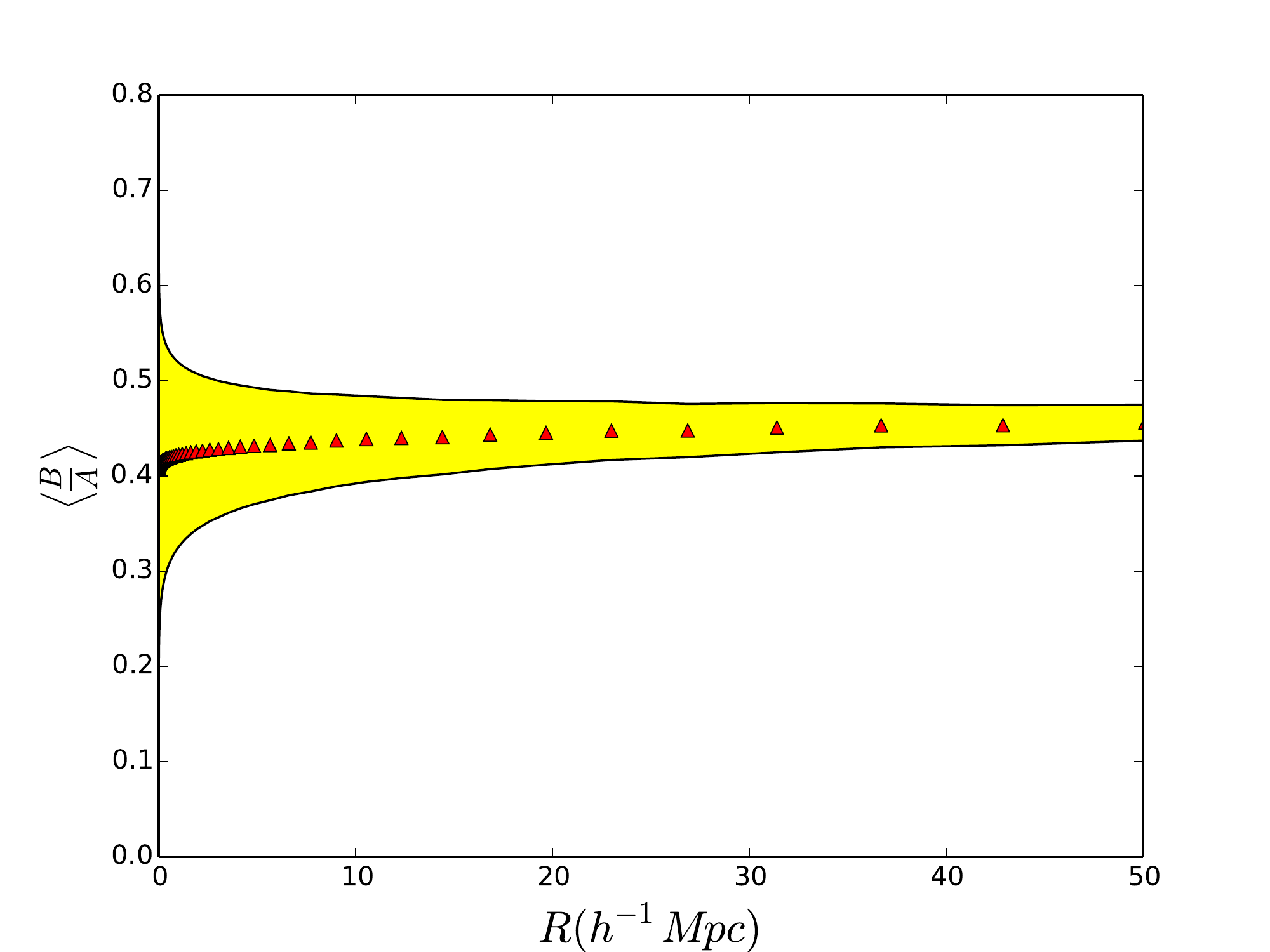}
\captionof{figure}{The average value of $\left\langle \frac{B}{A}\right\rangle$ and its dispersion, $\Delta B$ for various  radii (masses).}
\label{fig:ba}
\end{figure}

Fixing the virialization time, $t$, fixes $A$ (or the spherical collapse threshold) through  Equation (\ref{eq:A}), which in turn fixes the probability distribution of $B$ through Equation (\ref{eq:gaussian}):

\begin{equation}
\left\langle \frac{B}{A}\right\rangle =  \frac{\left\langle B \right\rangle}{A}
 = \frac{\left\langle AB \right\rangle}{\left\langle A^2 \right\rangle}
 \approx 0.42,
 \label{eq:AB}
\end{equation}
with a Gaussian dispersion from the mean given as,
\begin{equation}
\frac{\Delta B}{A} =  A^{-1} \sqrt{\frac{L}{\left\langle A^2 \right\rangle}} \approx  0.083\nu^{-0.6} ~~~~ ( 0.1 \leq \nu \leq 10)
%( This has same error as your paper)
,%\\
%\frac{\Delta B}{A} &\approx & 0.0052\nu^{-0.6} ( 0.1 \leq \nu \leq 10)%( This has smaller error because of the inclusion of the volume in A)
\label{eq:badisp}
\end{equation}
where $L = \left\langle A^2 \right\rangle\left\langle B^2 \right\rangle - \left\langle AB \right\rangle^2$, and $\nu \equiv \delta_{\rm sc}/\sigma(M)$ is the standard measure of peak height with $\delta_{\rm sc}  \approx 1.68 $. Fig \ref{fig:ba} shows the behaviour of $\left\langle \frac{B}{A}\right\rangle$ and its dispersion as a function of radius (mass). Note that, while Equations (\ref{eq:AB}-\ref{eq:badisp}) provide accurate fits for $\Lambda$CDM linear power spectrum, they can be used for arbitrary power spectra and cosmologies using their definitions in Sec. (\ref{sec:prob_ab}).

\begin{figure}
\centering
\includegraphics[width=1.0\columnwidth]{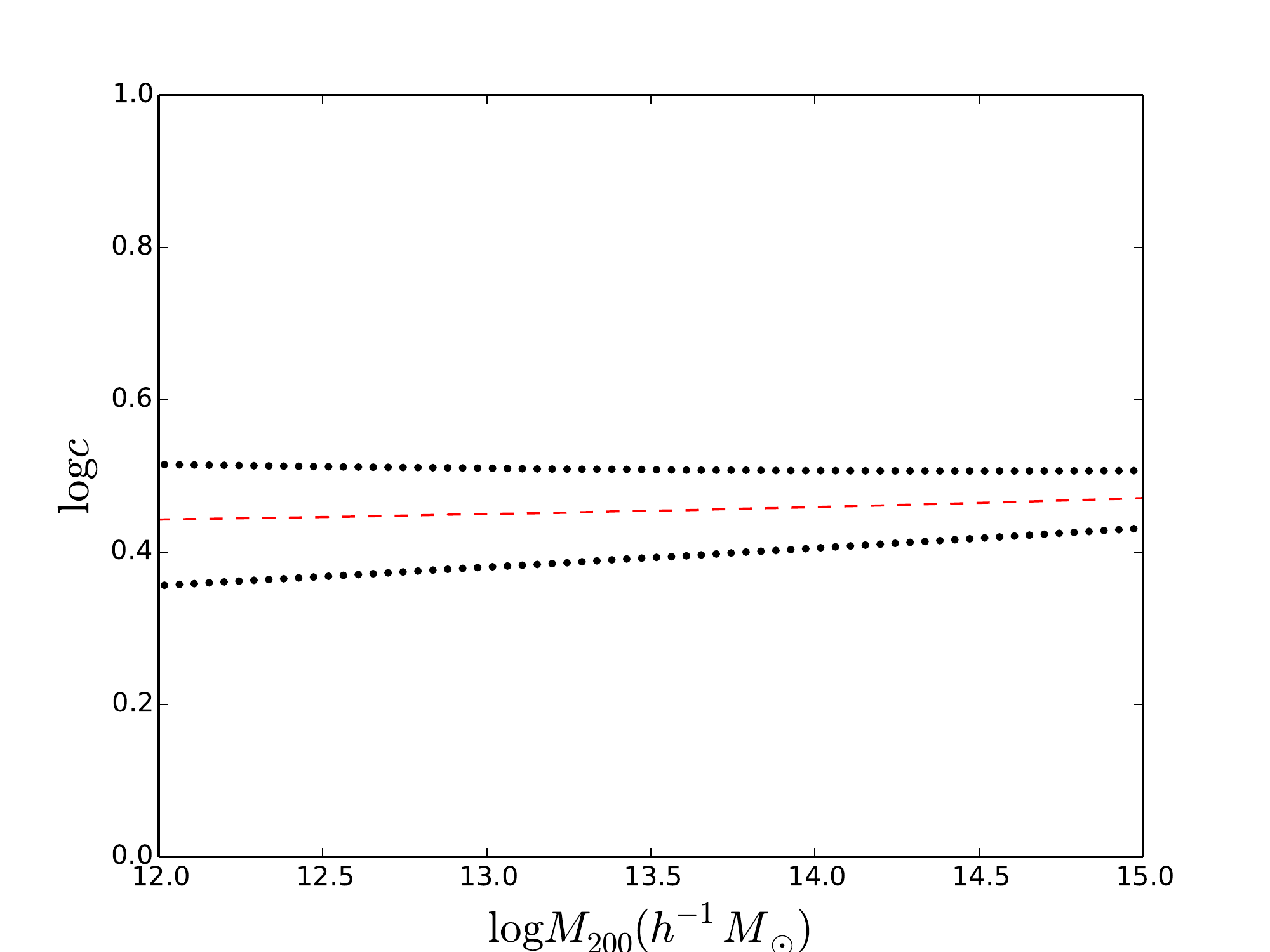}
\captionof{figure}{The mass--concentration relation of dark matter haloes with an NFW density profile derived from the spherical collapse model for the $\Lambda$CDM cosmology at $z$ = 0. The red dashed line gives the average value while the region between the thick black dots gives the dispersion in concentration for a fixed halo mass.}
\label{fig:ncandm}
\end{figure}

We are therefore completely armed with all the necessary tools to derive the concentration--mass relation, simply by assuming:
\begin{equation}
y_i = y_f
\end{equation}
in Equations (\ref{eq:c_einasto}-\ref{eq:badisp}). Fig. \ref{fig:ncandm} shows our derived relation for the spherical collapse model using the NFW profile. This shows a nearly constant relation, irrespective of mass, and a scatter that decreases with mass. For the Einasto density profile, the expected concentration is $10\%$ higher than that of the NFW profile. These results clearly do not agree with the well-known results from N-body simulations -- the concentration--mass relation decreases with mass \citep{b2,b3,b12,b38,b24,b29,b13}. However, as we will show in Section \ref{conclude}, they agree reasonably well at large masses. One reason for this is that the spherical collapse model is well suited for collapse of high mass haloes but fails at low masses \citep{b35}. The spherical collapse model also evolves weakly with redshift (our relation changes little with redshift through the $Ht$ variable $y_i$), thus agrees with the results of  \citet{b37} and \citet{b13} that concentration of high mass haloes evolves weakly with redshift.  In the next subsection, we incorporate the corrections due to the ellipticity of the low mass haloes to the mass--concentration relation.

\subsection{Ellipsoidal Collapse Model}
The spherical collapse model in its simplicity oversimplifies the formation of bound objects from collapse. Our modifications to the spherical collapse model stems from the assertion that perturbations in Gaussian density fields are triaxial \citep{c37}. Although the spherical collapse model makes reasonably simple analytic predictions regarding the shape of the mass function of bound objects,  when compared to simulations, it has more low mass haloes and less high mass haloes \citep{b35}. The considerable reduction of this discrepancy with haloes remodelled with the ellipsoidal collapse model motivates the consideration of a similar remodelled collapse in the concentration--mass relation. The spherical collapse model described in Section \ref{subsec:sphere} assumes that collapse occurs if the mean initial density of a region exceeds a critical value, $\delta_{\rm sc}$. This critical value is independent of mass or radius (only dependent on redshift, z) of the region and is thus known as the constant barrier. This implies that at a fixed redshift, all haloes with average initial overdensity greater than $\delta_{\rm sc}$  will collapse. However, \citet{b35} modified this relation for assumptions of ellipsoidal collapse, also known as the moving barrier. This modification is based on the fact that the critical overdensity for ellipsoidal collapse, $\delta_{\rm ec}$ depends on the mass or size of the collapsing region. An interesting consequence of this mass dependence is that smaller objects, which are more likely to be influenced by external tides, should have larger initial overdensities to hold them together as they collapse. This effect leads to a higher collapse time which verifies the results of \citep{c11}. Although a fixed mass fixes the collapse time (equation \ref{eq:A}) for the spherical collapse model, due to the range of ellipticities and prolatenesses in an ellipsoidal collapse, there is a range of collapse times for any fixed region (mass). 
 
\begin{figure*}
\includegraphics[width=1.\textwidth]{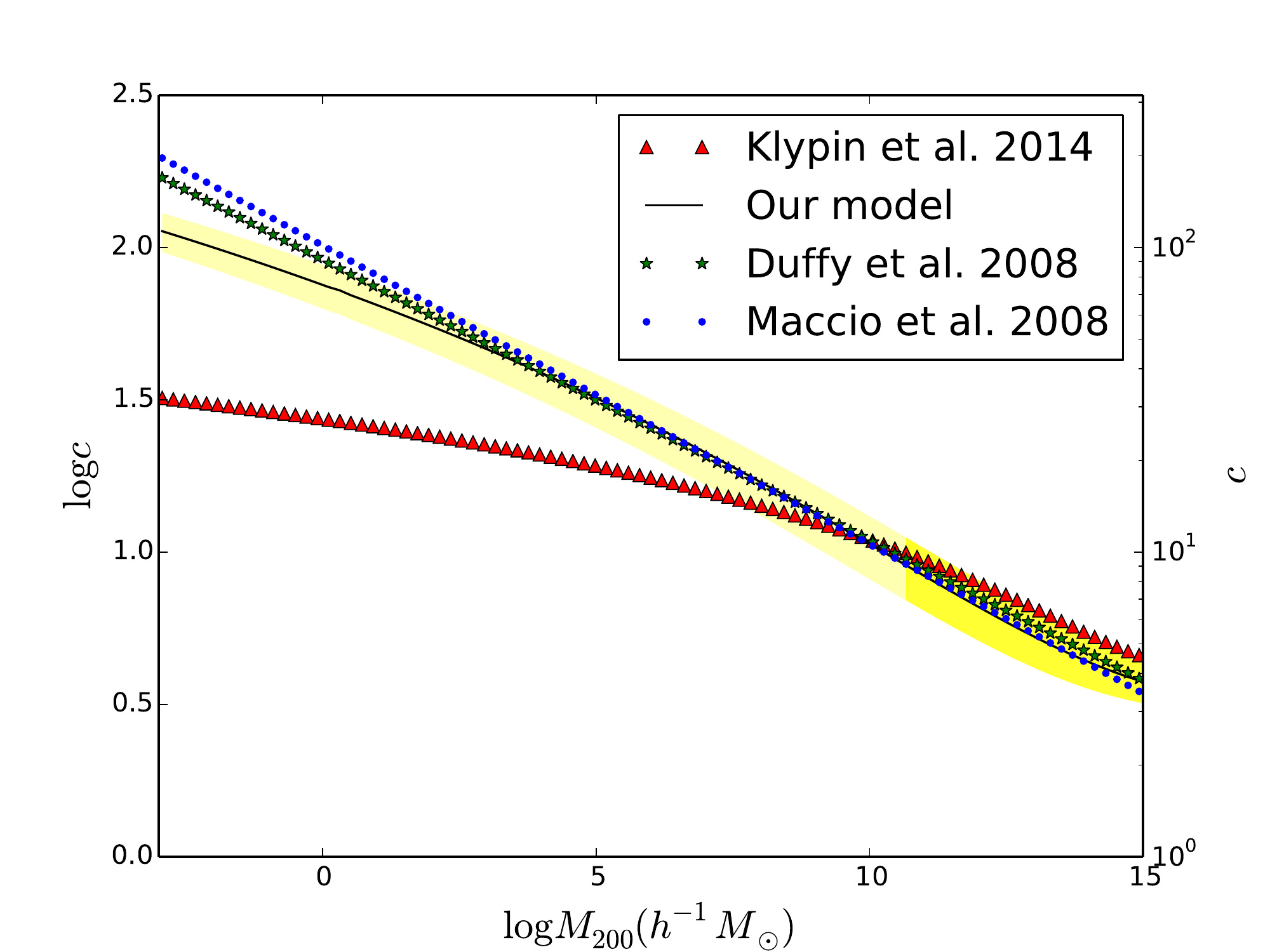}
\caption{
The concentration of dark matter haloes as a function of mass for haloes with NFW density profile and $\Lambda$CDM universe (WMAP5 Cosmology) at $z$ = 0. The black line shows the results of the mean concentrations derived from our model with the yellow region as the dispersion for fixed masses. Also shown are the concentrations from recent literature \protect \citep{c3,b15,c1}. The region in darker yellow show the range of masses probed by most N-body simulations, while lower masses use extrapolations by different groups. The decrease in concentration with mass can be interpreted as a result of the decrease in the critical collapse density as the mass increases. Previous results \protect \citep{b2,b3,b37} reveal that the halo concentration is a  measure of the density of the universe at formation since smaller masses form earlier.}
\label{fig:comcm}
\end{figure*} 
 
For the ellipsoidal collapse, the initial velocity field (related to the gravitational potential) is preferred relative to the initial density field. The Zeldovich type approximations \citep[see][for details]{b36} follow the perturbations in particle trajectories with the particle positions in terms of the Eulerian and Lagrangian coordinates. In a smooth universe with uniform density $\rho_b(t)$, the actual position of any particle $r(t)$ is related to its initial Lagrangian position, $q$ by 
\begin{equation}
 {\bf r}(t) = a(t){\bf q},
\end{equation}
while in the presence of growing density perturbations, we have
\begin{eqnarray}
{\bf r}(t) &=& a(t) {\bf x}(t) \nonumber\\
&=& a(t)[{\bf q} + D(t){\bf p}({\bf q})],
\end{eqnarray}
where $a(t)$ is the cosmic expansion scale factor, $D(t)$ is the linear growth rate and ${\bf p}({\bf q})$ is the velocity term. The conservation of mass implies that
\begin{eqnarray}
\rho({\bf r},t) &=& \frac{\rho_b(t)}{\det\left(\partial x_j/\partial q_i\right)}\nonumber \\
&=& \frac{\rho_b(t)}{\det\left[\delta_{ij} + D(t)(\partial p_j/\partial q_i)\right]}.
\end{eqnarray}
The tensor $\frac{\partial x_i}{\partial q_j}$ is known as the deformation tensor. This matrix can be diagonalized at every point, ${\bf q}$ to yield a set of eigenvalues as a function of ${\bf q}$. The eigenvalues, $\lambda_1$ $\geq$ $\lambda_2$ $\geq$ $\lambda_3$, define a coordinate system in which a certain volume preserves its original orientation upon deformation. Alternatively, one can also describe the shape of a region by its ellipticity, $e$, and prolateness, $p$, \citep{b35} defined by
\begin{equation}
e = \frac{\lambda_1 - \lambda_3}{2\delta} \makebox{	  and  	}  p = \frac{\lambda_1 + \lambda_3 - 2\lambda_2}{2\delta}
\end{equation}
The eigenvalues are ordered such that $e \geq 0$ if $\delta > 0$, and $-e \leq p \leq e$. A spherical region has $e = 0$ and $p = 0$. 
Following \citet{b35}, the evolution of an ellipsoidal perturbation is specified by the eigenvalues of the deformation tensor or alternatively by the density contrast $\delta$ and the initial ellipticity, $e$ and prolateness, $p$ of the linear tidal field. One can then construct the initial overdensity for collapse $\delta_{\rm ec}(e,p)$, for any e and p. An average collapse overdensity $\delta_{\rm ec}(\sigma)$, can be estimated on a scale $R$ parametrized by $\sigma$ by averaging over the distribution of $e$, $p$, and $\delta$. For $p$ = 0, the relation between $\delta_{\rm ec}$ and mass is fitted by 
\begin{equation}
\delta_{\rm ec}(\nu) \approx \delta_{\rm sc} \left(1 + \kappa \nu^{-2\gamma} \right),~~ {\rm with} ~\kappa= 0.47~{\rm and} ~\gamma= 0.615,
\label{eq:ec}
\end{equation}
where $\nu \equiv \delta_{\rm sc}/\sigma(M)$ is the standard definition of peak height.

\begin{figure*}
 \centering
\begin{subfigure}{1.0\columnwidth}
\centering
\includegraphics[width=1.\columnwidth]{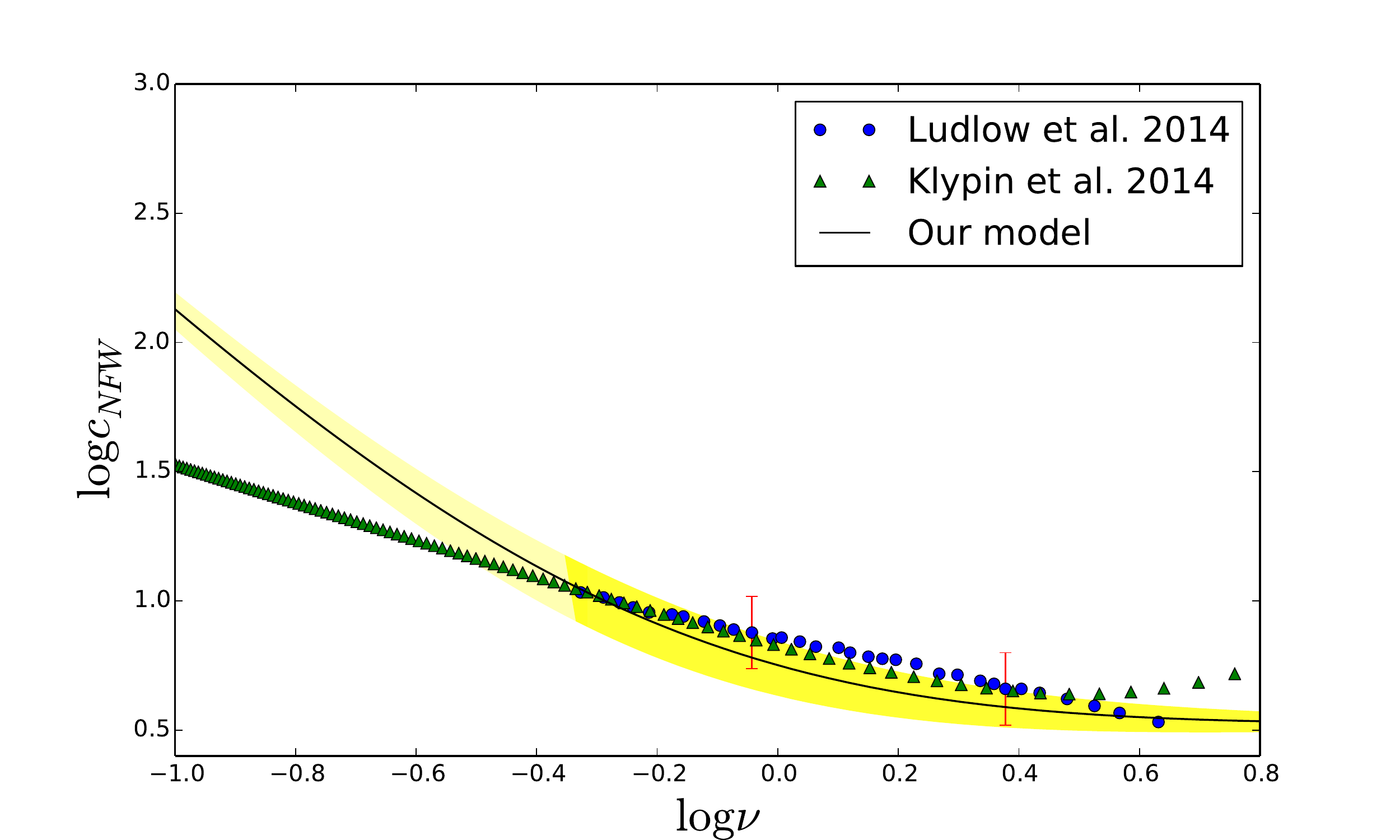}
\caption{}
\label{fig:newc}
\end{subfigure}%
\begin{subfigure}{1.0\columnwidth}
\centering
\includegraphics[width=1.0\columnwidth]{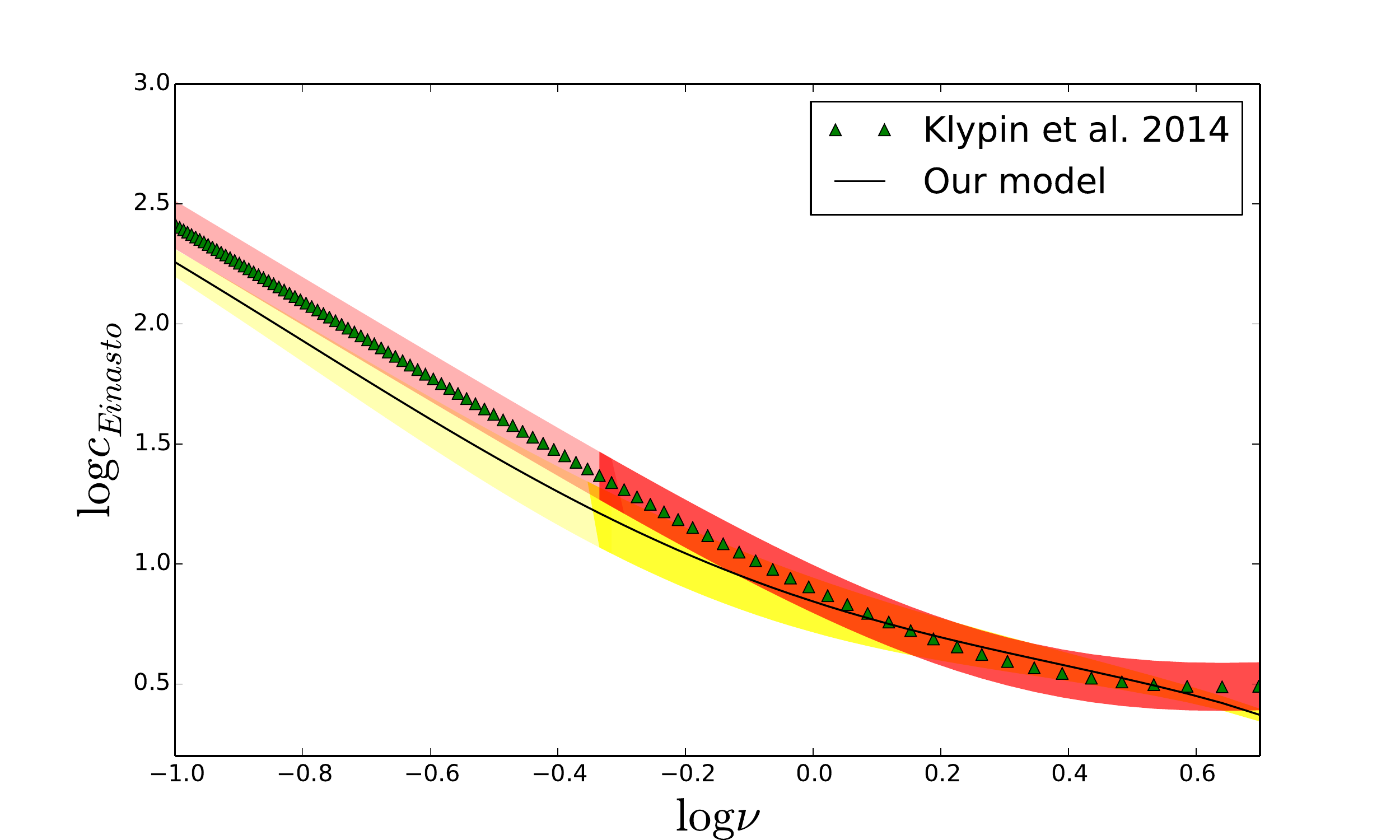}
\caption{}
\label{fig:allin1e}
\end{subfigure}%
\caption{
\textit{Left:} The concentration of dark matter haloes as a function of $\nu = \delta_c/\sigma(M,z)$ for haloes with NFW density profile for the  $\Lambda$CDM universe (Planck Cosmology) at $z$ = 0. The parametrization in terms of $\nu$ is preferable since it incorporates both the mass dependence and redshift dependence of the concentration.  As in Figure \ref{fig:comcm}, the black line shows the results of the concentrations derived from our model, while the regions in darker colour show the mass range probed by N-body simulations. Also shown are the concentrations from recent literature \protect \citep{b13,c1}. The red error bars show the dispersion of the results for \protect \citet{b13}. \textit{Right:} The concentration of dark matter haloes as a function of $\nu = \delta_c/\sigma(M,z)$ for haloes with Einasto density profile for the  $\Lambda$CDM universe (Planck Cosmology) at $z$ = 0. As in \ref{fig:comcm}, the black line (yellow region) shows the results (dispersions) of the concentrations derived from our model. Also shown are the concentrations from \protect \citet{c1} with its $10\%$ dispersion at fixed mass. }
%\label{fig:newc}
\end{figure*}

An important feature of Equation \ref{eq:ec} is that the critical overdensity is larger for less massive systems since $\sigma(M)$ decreases for massive systems, thus small mass haloes were more likely to be formed from ellipsoidal collapse. This was subsequently corroborated by later studies, e.g., \citet{b40}, who investigated the protohaloes of collapsed haloes in simulations. Equation \ref{eq:ec} is extremely useful since it allows us to incorporate the effect of ellipsoidal collapse into our concentration--mass relation. Redshift effects and cosmology dependence are robustly incorporated through the model dependent $\delta_{\rm sc}$ and the model dependent power spectrum, $\sigma^2(M)$. The concentration--mass relation for ellipsoidal collapse is only modified through $y_i$ since that is the bit of the relation related to the initial state of the region before collapse. Thus our new relation becomes
\begin{equation}
(Ht)^{2/3} y_{\rm ec} = \left\langle \frac{B}{A} \right\rangle \left(1 + \kappa \nu^{-2\gamma} \right) \pm  \frac{\Delta B}{A},
\label{eq:yec}
\end{equation}
which combined with equations (\ref{eq:c_einasto}-\ref{eq:c_nfw}, \ref{eq:AB}-\ref{eq:badisp}) and assuming $y_{\rm ec} =y_i=y_f$, fixes the concentration--mass relation. It is easy to see that the modified relation recovers the scenario for the spherical collapse model at high masses.

It is interesting to understand why the ellipsoidal collapse {\it only} affects the mean, but not the dispersion of $y \propto E_i \propto B$. Since $B$ is a random Gaussian variable that follows the distribution (\ref{eq:gaussian}), its dispersion is fixed by the moments of power spectrum for a given size/mass of the spherical region, independent of the time of collapse. However, its mean is set by its cross-correlation with $A$, which is proportional to the linear collapse threshold. Therefore, just boosting the linear collapse threshold through Equation (\ref{eq:ec}) boosts the mean of $y\propto B$ by the same factor, but does {\it not affect} its dispersion.

The predicted concentration from this model is shown in Fig. \ref{fig:comcm}. The black line depicts the mean concentration as a function of mass for our model, with the yellow region as the $68\%$ dispersion from the mean value. Our model shows a decrease in the concentration with increasing mass and asymptotes to a constant value for large masses, consistent with simulation results.

An alternative common parametrization of concentration is in terms of the dimensionless peak height parameter, $\nu$, which captures the bulk of cosmology dependence (as is evident in our analytic fits)
Fig \ref{fig:newc} and Fig \ref{fig:allin1e} show the concentration of haloes as a function of $\nu$ for the NFW profile and the Einasto profile, respectively. For the Einasto profile, we chose the shape parameter $\alpha$ for relaxed haloes given by \citet{c1}, 
\begin{equation}
\alpha = 0.115 + 0.014\nu^2.
\label{eq:al}
\end{equation}

\subsection{Is total energy conserved?}

While we have included the effect of external tidal fields on the ellipsoidal collapse time above, we have ignored its effect on the conservation of total energy. Furthermore, the ejection of 5-15\% of particles during halo mergers (primarily from the minor halo) may lead to a systematic bias in our estimated total energies  \citep{2009MNRAS.397..775J,c10}. However, the close agreement between our results and those found in N-body simulations (e.g., Figs \ref{fig:comcm}, \ref{fig:newc}, \ref{fig:allin1e}, and \ref{fig:zsign}) suggests these effects may be negligible for total energy estimates that go into our predictions. Some possible reasons are:

\begin{enumerate}

\item The collapse time depends  on the energy conservation for the last infalling shell, which is more susceptible to tidal effects than the random particle inside the halo.  

\item The delay in ellipsoidal collapse is possibly due to tidal fields; infalling particles do not follow straight lines, and thus density doesn't grow as fast as it would in spherical collapse. This process doesn't require the tidal field doing work on the particles. For example, a static tidal field doesn't change the energy but delays the collapse. Similarly, the work done by tidal field averages to zero for a spherical shell. 

\item The ejected particles in mergers have close to zero energy and/or mostly fall back into the halo in subsequent mergers.
\end{enumerate}

%\textcolor{red}{ Furthermore, the ejection of 5\%-15\% of particles during halo mergers (primarily from the minor halo) may lead to a systematic bias in our estimated total energies. It is worthy to note that \citet{c10} pointed out that about $5-15\%$ particles are ejected from the minor halo during minor mergers while the particles in the major halo are basically conserved. This finding also corroborates our results that the spherical collapse works better for larger halos (whose energy is basically conserved) than for smaller halos. The loss of particles observed by \citet{c10} leads to a gain in the total energy of the minor halo during mergers. The correspondence between our results and simulations suggests that this effect is minimal relative to the change in collapse time (discussed above) for the ellipsoidal collapse model. Furthermore, any change in the total energy for the smaller halos seems to be smaller than the dispersion in concentrations at fixed mass derived from $\frac{\Delta B}{A}$.}

Therefore, we conclude that the assumption of total energy conservation might be a reasonable approximation for ellipsoidal collapse. 

\section{Halo Velocity Dispersion}\label{sec:sigma}
The halo velocity dispersion is a property of haloes related to the final energy of the halo. The spherical collapse model \citep{b26}  predicts the scaling relation of the one dimensional halo velocity dispersion in terms of halo mass $M_{200}$ and the dimensionless Hubble parameter, $h(z) \equiv H(z)/[100 ~{\rm km/s/Mpc}]$ as
\begin{equation}
\sigma_{1d} = \sigma_{15}\left( \frac{M_{200} }{10^{15} M_{\odot}/h(z)} \right)^{1/3}. 
\end{equation}
While $\sigma_{15}$ is a constant in the original top-hat collapse model, inclusion of ellipsoidal effects and surface terms in virial theorem lead to a prediction of the normalization $\sigma_{15}$ which depend on $y$ and $\omega$ through Equations \ref{eq:mu}-\ref{eq:yn}:
\begin{eqnarray}
\sigma_{15}  &=& \left(2 \pi G H(z)\times 10^{15}  M_{\odot} \over h(z)\right)^{1/3}\sqrt{\frac{y(1 + \omega)}{2(1 - \omega)}} \nonumber\\
&=& (1394 ~{\rm km/s}) \sqrt{\frac{y(1 + \omega)}{2(1 - \omega)}},
\label{eq:sig}
\end{eqnarray}
where we used:
\begin{equation}
K=\frac{3}{2} M \sigma_{1d}^2,
\end{equation} 
to relate 1-D (mass-weighted) velocity dispersion to the total kinetic energy.

\begin{figure*}
 \centering
\begin{subfigure}{1.0\columnwidth}
\includegraphics[width=1.0\columnwidth]{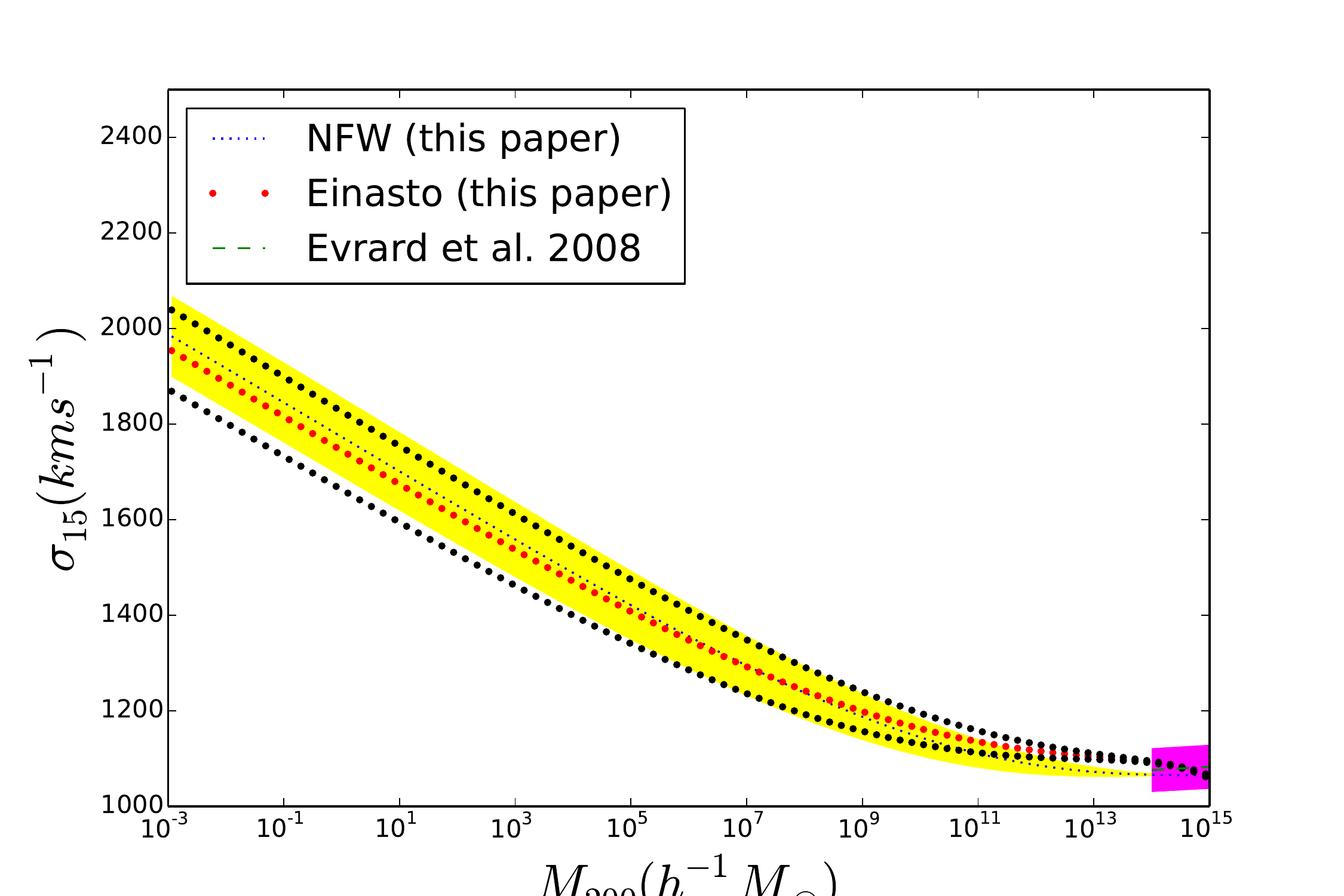}
\caption{}
\label{fig:sign}
\end{subfigure}%
\begin{subfigure}{1.0\columnwidth}
\includegraphics[width=1.0\columnwidth]{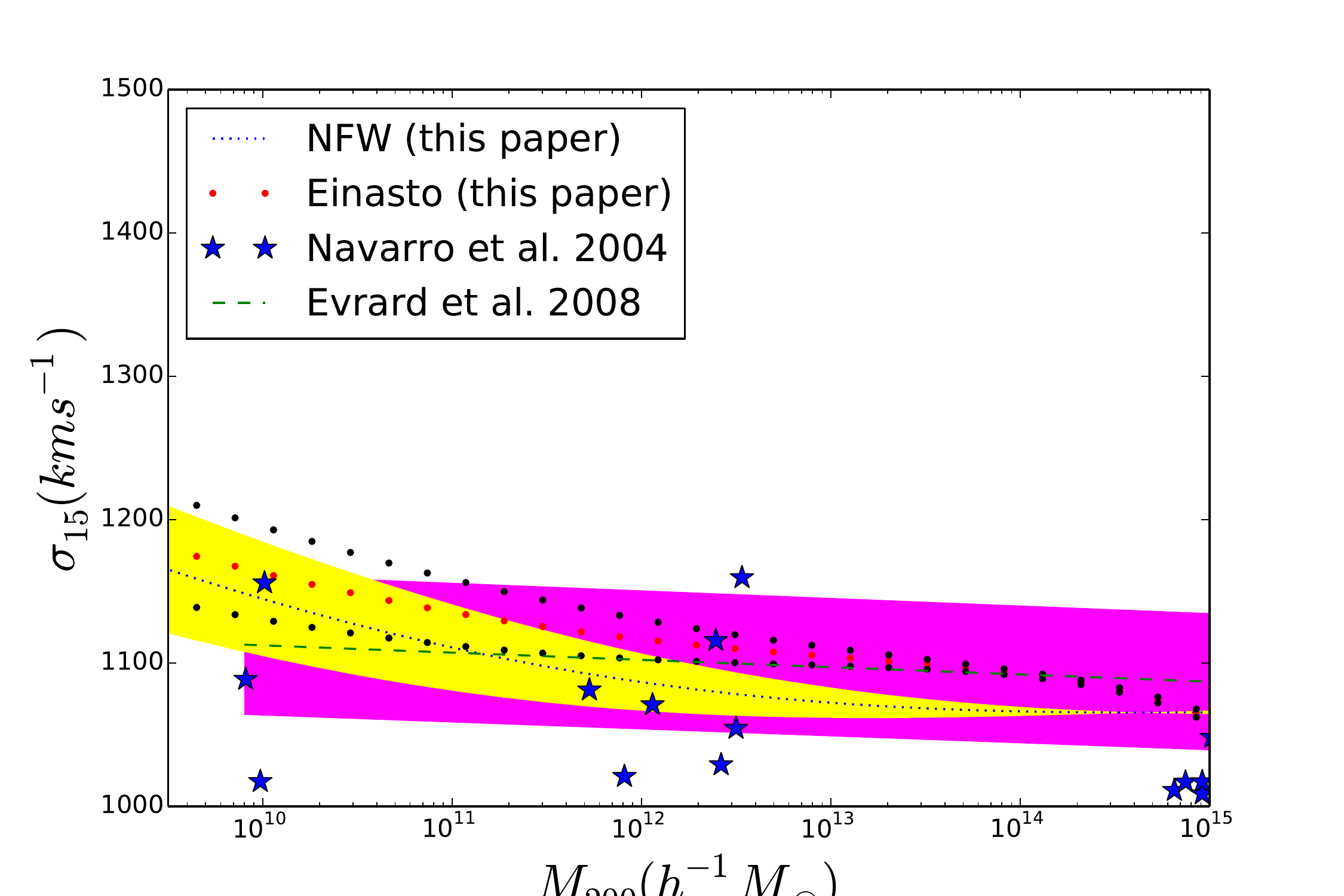}
\caption{}
\label{fig:zsign}
\end{subfigure}%
\caption{\textit{Left:} The normalization, $\sigma_{15} \equiv \sigma_{1d}{\left( \frac{M_{200} }{10^{15} M_{\odot}/h(z)} \right)^{-1/3}}$ of the one dimensional velocity in terms of the mass of the halo. The blue dashed lines (red dots) are our prediction for the NFW (Einasto) density profile. The yellow region (region between the black dots) are the dispersion at fixed masses for the NFW (Einasto) profile. A comparison (green dashed line) is made with the results of simulations in \protect \citet{c5} and the dispersion at fixed mass given by the magenta region. \textit{Right:} A close-up plot of simulated regions ($10^{10}h^{-1}M_{\odot} \leq M_{200} \leq 10^{15}h^{-1}M_{\odot}$) from the plot on the left. Markers and colours on the plot are same with the description on the left. Blue stars are the plots from the simulation of \protect \cite{c34} shown in Fig 6 of \protect \cite{c5}.}
\label{}
\end{figure*}
\begin{figure}
 \centering
\includegraphics[width=1.0\columnwidth]{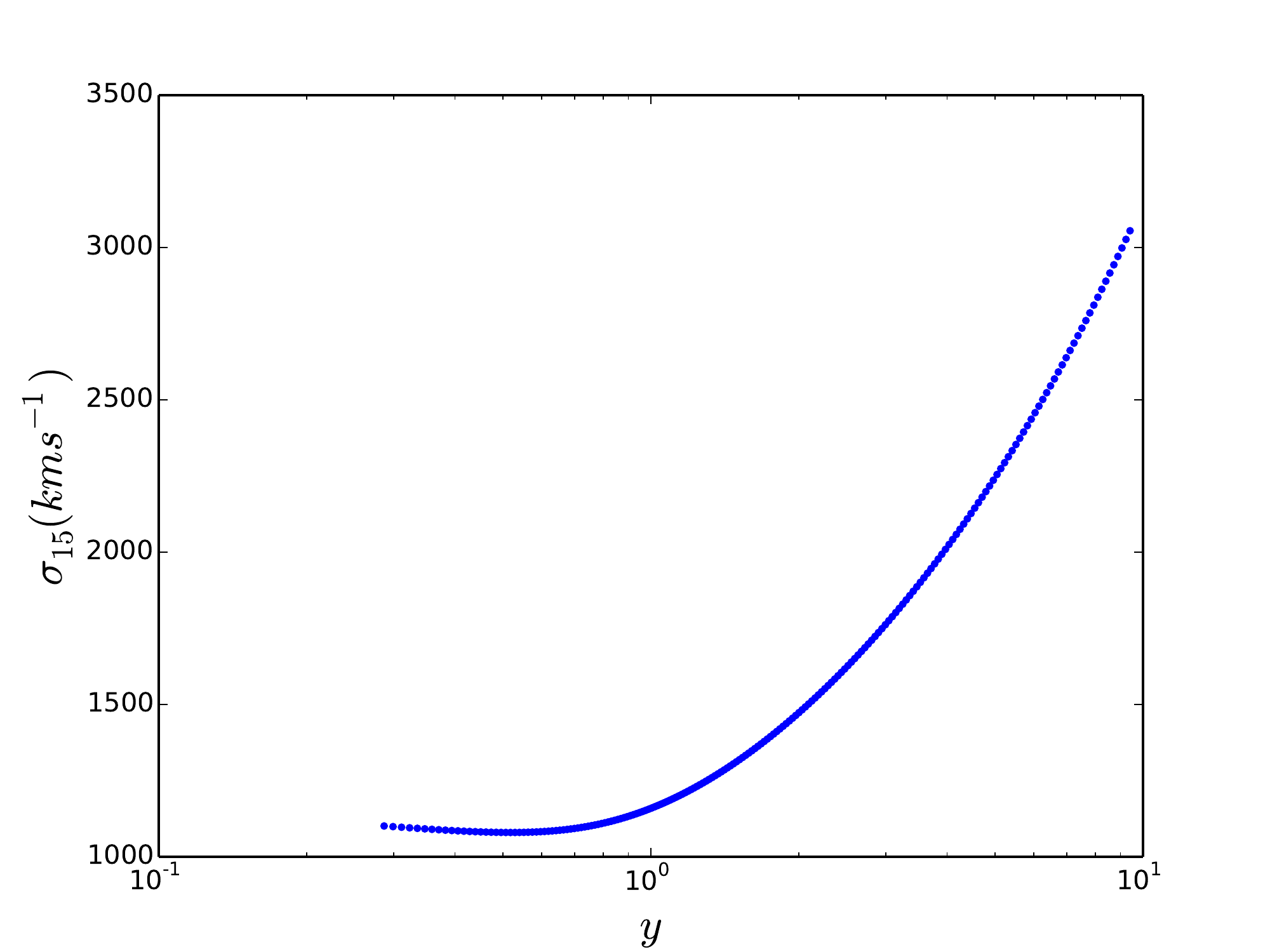}
\caption{The normalization of the one dimensional velocity dispersion,$\sigma_{15} \equiv \sigma_{1d}{\left( \frac{M_{200} }{10^{15} M_{\odot}/h(z)} \right)^{-1/3}}$, as a function of the dimensionless total energy parameter, $y$, for the NFW profile at $z$ = 0. We see that $\sigma_{15}$ reaches a minimum at $y \simeq 0.54$, which means velocity dispersion becomes insensitive to the initial/total energy at fixed mass. This explains the low scatter in velocity dispersion of massive simulated haloes. }
\label{fig:sigy}
\end{figure}

The dependence of Equation \ref{eq:sig} on the concentration (through $y$ and $\omega$) and the shape parameter $\alpha$ (for the Einasto profile) yields a mass dependence, if energy is conserved. This relation (shown in Figure \ref{fig:sigy}), for the NFW profile is well fit (to $1\%$ precision) by:
\begin{equation}
\sigma_{15}({\rm km/s}) \simeq 241 \ln{y}^2 + 295 \ln{y} + 1156,
\label{eq:s_15_nfw}
\end{equation}
and 
\begin{equation}
\sigma_{15}({\rm km/s}) \simeq b_1(\alpha)\ln(y)^3 + b_2(\alpha)\ln(y)^2 + b_3(\alpha)\ln(y) +b_4(\alpha),
\label{eq:s_15_einasto}
\end{equation}
for the Einasto density profile (at $1\% ~(10\%)$ precision for $\alpha <0.25 ~(0.5)$ ). The $b_i$'s are second order polynomials of $\alpha$. For $0.1 < $ $\alpha $ $ < 0.52$, they are given as 
\begin{eqnarray}
b_1(\alpha) &=& -173 \alpha^2 + 237 \alpha - 14,\\ \nonumber
b_2(\alpha) &=& -389 \alpha^2 - 378 \alpha + 287,\\ \nonumber
b_3(\alpha) &=& 1540 \alpha^2 - 195 \alpha + 244,\\ \nonumber
b_4(\alpha) &=& 71 \alpha^2 - 287 \alpha +1205,
\end{eqnarray}
while $\alpha$ is given by equation \ref{eq:al} for the Einasto profile, and $y$ by equation \ref{eq:yec}. The resulting normalization, $\sigma_{15}$, is shown as a function of mass in Fig \ref{fig:sign}. Also shown are the mean value and the scatter of the normalization: 
\begin{equation}
\sigma_{15}({\rm km/s}) = 1083 \pm 46, ~~{\rm (Evrard~ et~ al.~ 2008)}
\label{eq:evrard}
\end{equation}
 from the simulations of \citet{c5}, for cluster mass haloes in the range $10^{14} h^{-1} M_{\odot}$ - $10^{15} h^{-1} M_{\odot}$. Our predictions of the normalization for both the NFW and Einasto profiles agree very well with their results within the dispersion. Fig \ref{fig:zsign} zooms in on expected normalization for lower mass haloes. Also shown are the results from the simulation of \cite{c34}.

The robustness and small scatter (or cosmology independence) of our predictions for velocity dispersion of massive haloes, can be easily understood as the balance between the surface pressure and self-gravity of the haloes. Isolated self-gravitating systems have negative heat capacity, while virialized objects bound by surface pressure (such as air in a room) have positive heat capacity. Haloes more massive than $\sim 10^{14} M_\odot$ ($y < 0.54$)  have low enough concentration that the surface pressure drives the sign of heat capacity to positive values (Figure \ref{fig:sigy}). Therefore, cluster mass haloes have inverse heat capacity $\sim 0$, which means that their temperature (or velocity dispersion) is insensitive to their energy, and thus initial conditions or cosmology. In other words, $\sigma_{15}$, for e.g. NFW profile (Equation \ref{eq:s_15_nfw}), has a minimum at:
\begin{equation}
\sigma_{15}|_{\rm min} \simeq 1066~ {\rm km/s},
\end{equation}
which is well within the simulated range (\ref{eq:evrard}) and consistent with the positive skewness seen in \citet{c5}.

\begin{figure*}
 %\centering
%\begin{subfigure}{1.0\columnwidth}
%\centering
\includegraphics[width=0.95\textwidth]{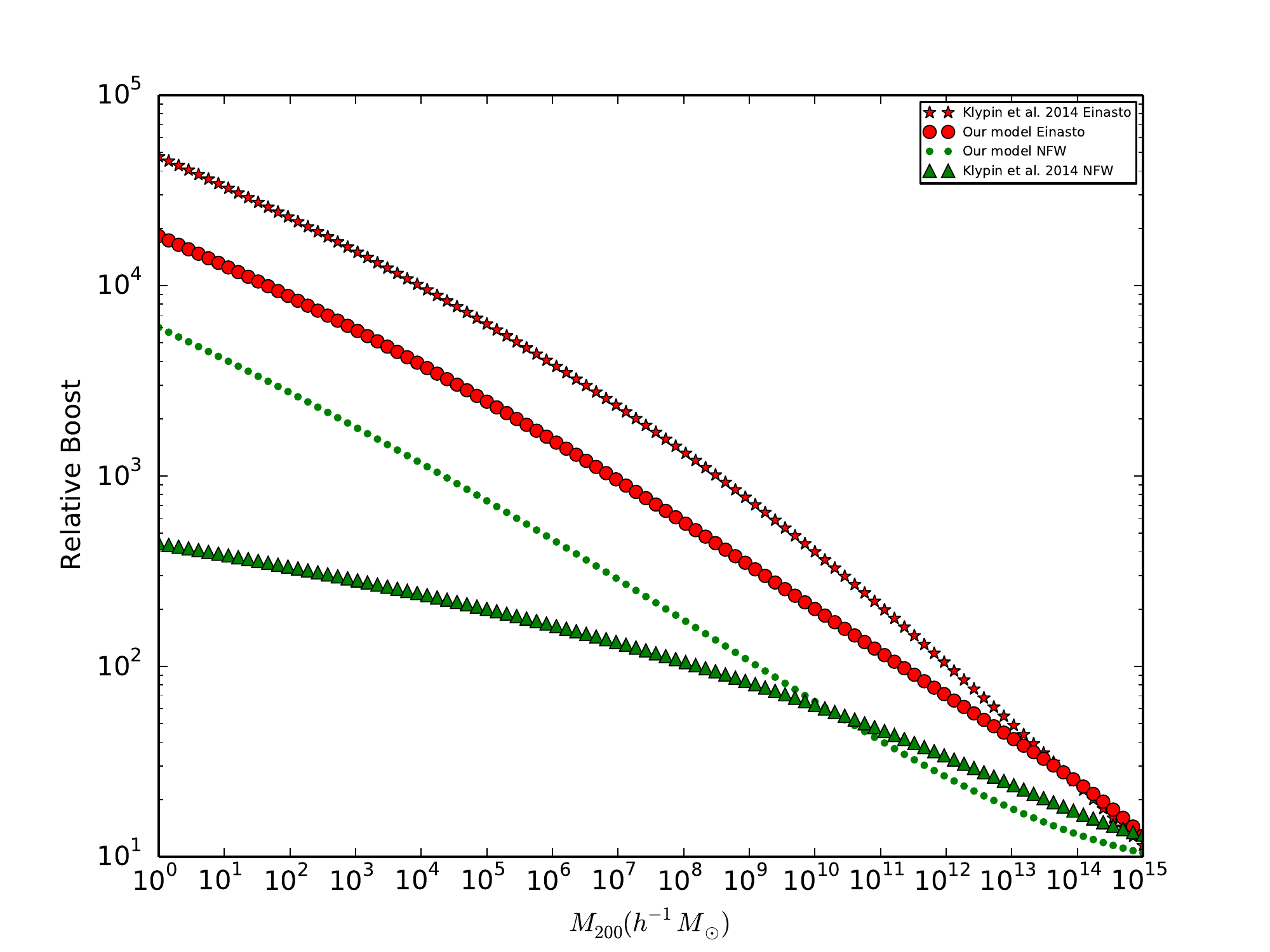}
\caption{
The expected annihilation boost from a halo with the NFW or Einasto profile, relative to that of a uniform density halo at $z = 0$. The predictions on the plot are based on our model and those of \protect \citet{c1}. The low mass haloes are expected to dominate the boost from dark matter annihilation.}
\label{fig:boost}
\end{figure*}
\section{Discussion and Conclusion}\label{conclude}
Qualitatively, the concentration of dark matter haloes is a decreasing function of mass  that flattens at very low and very high masses. This feature -- decrease of concentration with mass -- can be attributed to major mergers that lead to bigger haloes, but disrupt the inner regions of haloes that subsequently decreases concentration. We have used the conservation of energy to derive the concentration--mass relation of CDM haloes. The robustness of our prescription lies in the fact that one can compute the concentration--mass relation for any cosmology by using the cosmology-dependent power spectrum, cosmic age, and linear ellipsoidal collapse threshold, $\delta_{\rm ec}.$ Our results show that the concentration of a halo is set by the initial total energy of the region prior to collapse, as well as  the cosmological parameters at collapse time. Small mass haloes mostly collapse from ellipsoidal regions and are better described by the ellipsoidal collapse model while large mass haloes are well described by spherical collapse model. %This is largely due to the fact that large mass haloes collapse later when the nearby small mass haloes, which collapsed earlier, have become spherical \citep{b15} and thus can be well described by spherical collapse. 
%Our results also corroborate with those of \citet{b40} -- the protohaloes of small masses are mostly ellipsoidal while those of large mass haloes are mostly spherical. {\color{red} {\bf [The last two sentences don't make sense.]}}.

Several analytical relations exist in the literature for the concentration--mass of CDM haloes, though fitted through numerical simulations. \citet{b13}'s model generates the concentration of CDM haloes from simulated and analytic mass accretion histories of haloes. Their model predicts that at very low masses, the concentration varies slower than one should expect from power law fits at high masses. This feature, according to the authors,  is a consequence of the shallow slope of the linear power spectrum at very low masses/small scales. However, at high masses, it approaches a constant value. This characteristic corroborates the expectations from our model and is consistent with earlier results of \citet{b41}. \citet{c12} also predict a theoretical mass-concentration relation through their theoretical density profile which relates to the accretion rate of halos. The free parameter of their model is then fit to the mass-concentration relation of \citet{b41}. The concentration-mass relation agrees with ours within the simulated range, flattening around $10^{15}M_{\odot}$ with a surprising upturn beyond $2 \times 10^{15}M_{\odot}$. Although the concentrations derived from our model may be different when compared to those from the millennium simulations at some masses, it is marginally consistent within the range of dispersion as shown in Fig. \ref{fig:newc}. Our predicted halo concentrations also have dispersions around the median at fixed masses \citep{b12,b39}. A novel feature of our prediction is the decrease in the dispersion of the concentration with mass. This agrees with the results of \citet{b38}, which suggests it may be the result of massive haloes collapsing recently and are thus more homogeneous. On average for different mass bins, \citet{b38} had dispersions in their concentrations of about $\sigma_{log_{10}c} = 0.1$ for relaxed haloes, which is marginally consistent with our dispersions at fixed a masses for medium-sized haloes. At $z = 0$, our predictions are consistent with the concentrations of \citet{c23} and \citet{c39} for very small microhaloes. Our method does not suffer from the lack of scalability \citep{b29}, it is therefore applicable to any set of cosmological parameters. Though our results have been exclusively reported for $z = 0$, it is applicable to different redshifts through the redshift dependence of $\nu$ -- the usual scaling of the variance $\sigma(m)$ by the cosmological growth function, $D(z)$ and slightly through the $Ht$ parameter in the definition of $y$, Equation \ref{eq:yemp}.

As we noted in Section \ref{sec:intro}, the concentration--mass relation plays a crucial role in the expected boost from the annihilation of dark matter haloes. The expected boost to dark matter annihilation signal from a halo is proportional to the square of its density profile. In Fig. \ref{fig:boost}, we compare the expected boost from a halo with an NFW or Einasto profile relative to that of halo of uniform density of the same mass and radius. The comparison highlights the fact that different functional forms that fit simulated haloes rather well (e.g. NFW v.s. Einasto) could yield vastly different predictions, when extrapolated down to the bottom of CDM hierarchy \citep[a point also recently made in][]{2015arXiv150802713Z}. Therefore, as we see in Fig. \ref{fig:boost}, a physical prescription such as energy conservation which have adopted here can lead to much more robust predictions (even though it is still limited by the uncertainty in the profile).  

In summary, to derive the dependence of concentration and velocity dispersion of haloes on their mass for a parametric halo profile, one should follow the following prescription:
\begin{enumerate}
\item Derive the initial energy of a halo in terms of its initial radius (mass) and the random Gaussian parameters (density perturbations)($y_i$, \ref{eq:AB}, with dispersion \ref{eq:badisp}).
\item{Correct for initial asphericities (\ref{eq:yec})}
\item Relate the final energy of the halo to the concentration of the halo. In principle, one should include the existence of a non zero pressure at the boundary of the halo which incorporates the velocity anisotropy profile of the halo (Equation \ref{eq:c_einasto}-\ref{eq:c_nfw}).
\item Express the concentration of the halo in terms of the mass or the dimensionless peak height parameter, $\nu$ or mass.
\item Same energetic arguments give 1-D velocity dispersions in terms of $y_i$ through Equations (\ref{eq:s_15_nfw}-\ref{eq:s_15_einasto}).
\end{enumerate}

And here is some good news for the impatient reader! All these steps can be combined into three lines to give concentration at 10\% precision, for arbitrary redshift, cosmology, and halo mass (as long as the power spectrum does not deviate too widely from $\Lambda$CDM):

\begin{empheq}[box=\fbox]{align}
y =  \frac{0.42+ 0.20 \nu^{-1.23}\pm 0.083 \nu^{-0.6}}{(Ht)^{2/3}}, 
\label{eq:yemp}\\
%\log{c_{\rm Einasto}} &=& a_1(\alpha) \log{y_f}^2 + a_2(\alpha) \log{y_f} + a_3(\alpha), \\
\log{c_{\rm NFW}} = 0.78 \ln{y} + 1.09.\\
\frac{\sigma_{1d,{\rm NFW}}({\rm km/s})}{\left( \frac{M_{200} }{10^{15} M_{\odot}/h(z)} \right)^{1/3}} = 241 \ln{y}^2 + 295 \ln{y} + 1156
\end{empheq}
Similar fits are also given for Einasto profile in Eqs. \ref{eq:c_einasto} and \ref{eq:s_15_einasto}, which depend on $\alpha$ (Recall that $\alpha$, which is dependent on $\nu$, is the shape parameter for the Einasto profile).

The effectiveness of our method lies in the fact that it utilizes the mass variance  (generally, the full power spectrum) rather than the traditional method of finding a cosmology-dependent power law relation between concentration and mass; thus it is suitable for all cosmological models and comparisons with simulations. We must note that our model is most reliable for relaxed {\it and} isolated haloes that are observed today. It may not successfully predict the concentration of subhaloes or haloes that recently merged because we have assumed virial equilibrium at the time of observation. Furthermore, as concentration--mass relations from different simulations have significant differences at fixed masses, one may not expect our predictions to precisely match results from any single group. However, we note that our predictions for concentration of Einasto haloes are in good agreement with both measurements {\em and} extrapolations from simulations (Fig. \ref{fig:sign}). A more realistic prescription for ellipsoidal collapse, which may incorporate the non-linear effects of the tidal field and the initial geometry of the collapsing region (on {\it both} collapse time {\it and} total energy conservation), should improve the accuracy of our predictions. This is an avenue for future exploration. 

\section*{Acknowlegdements}
We thank Gus Evrard, Matt McQuinn, James Taylor and Jesus Zavala for useful discussions and comments on this manuscript. We also thank the anonymous referee for helpful comments. Our research is supported from the Perimeter Institute for Theoretical Physics and the University of Waterloo. Research at the Perimeter Institute is supported in part by the Government of Canada through Industry Canada, and by the Province of Ontario through the Ministry of Research and Information (MRI).

\label{lastpage}

\begin{thebibliography}{99}
\bibitem[\protect\citeauthoryear{Afshordi \& Cen}{2002}]{b23} Afshordi N., Cen R., 2002, ApJ, 564, 669
\bibitem[\protect\citeauthoryear{Anderhalden \& Diemand}{2013}]{c23} Anderhalden D., Diemand J., 2013, JCAP, 04, 009
\bibitem[\protect\citeauthoryear{Binney \& Tremaine}{1987}]{b28} Binney J., Tremaine S., 1987, Princeton University Press
\bibitem[\protect\citeauthoryear{Bullock et al.}{2001}]{b12} Bullock J. S., Kollat T. S., Sigad Y., Sommerville R. S., Kravtsov A. V., Klypin A. A., Primack J. S., Dekel A., 2001, MNRAS, 321, 559
\bibitem[\protect\citeauthoryear{Carucci et al.}{2014}]{c10} Carucci I. P., Sparre M., Hansen S. H., Joyce M., 2014, JCAP, 06, 057
\bibitem[\protect\citeauthoryear{Clowe et al.}{2006}]{b4} Clowe D., Brada{\v c} M., Gonzalez A. H., Markevitch M., Randall S. W., Jones C., Zaritsky D.,  2006, ApJ, 648, 109
\bibitem[\protect\citeauthoryear{Cole \& Lacey}{1996}]{b34} Cole S., Lacey C., 1996, MNRAS, 281, 716
\bibitem[\protect\citeauthoryear{Dehnen \&  McLaughlin}{2005}]{b7} Dehnen W., McLaughlin D., 2005, MNRAS, 363, 1057
\bibitem[\protect\citeauthoryear{Del Popolo \&  Gambera}{1998}]{c11} Del Popolo A., Gambera M., 1998, A\& A, 337, 96
\bibitem[\protect\citeauthoryear{Doroshkevich}{1970}]{c37} Doroshkevich A. G., 1970, Astrofizika, 3, 175
\bibitem[\protect\citeauthoryear{Duffy}{2008}]{c3} Duffy A. R., Schaye J., Kay S. T., Dalla V. C., 2008, MNRAS, 390, 64
\bibitem[\protect\citeauthoryear{Dutton \& Macci\`o}{2014}]{c2} Dutton A. A., Macci\`o A. V., 2014, MNRAS, 441, 3359
\bibitem[\protect\citeauthoryear{Einasto}{1965}]{c36} Einasto, J., 1965, Trudy Astrofizicheskogo Institute Alma-Ata, 5, 87
\bibitem[\protect\citeauthoryear{Eke, Navarro \& Steinmetz}{2001}]{b33} Eke V. R., Navarro J. F., Steinmetz M., 2001, Apj, 554,114
\bibitem[\protect\citeauthoryear{Evrard et al.}{2008}]{c5} Evrard, A. E., Bialek, J., Busha, M., White, M., Habib, S., Heitmann, K., Warren, M., Rasia, E., Tormen, G., Moscardini, L., Power, C., Jenkins, A. R., Gao, L., Frenk, C. S., Springel, V., White, S. D. M., Diemand, J., 2008, Apj, 672,122
\bibitem[\protect\citeauthoryear{Gao et al.}{2008}]{b37} Gao L., Navarro J. F., Cole S., Frenk C. S., White S. D. M., Springel V., Jenkins A., Neto A. F., 2008, MNRAS, 387, 536
\bibitem[\protect\citeauthoryear{Gunn \& Gott}{1972}]{b26} Gunn J. E., Gott J. R. III, 1972, ApJ, 176
\bibitem[\protect\citeauthoryear{Hansen \& Moore}{2006}]{b17} Hansen S. H., Moore B., 2006, NewA, 11, 333
\bibitem[\protect\citeauthoryear{Hansen}{2009}]{b31} Hansen S. H., 2009, ApJ, 694, 1250
\bibitem[\protect\citeauthoryear{Henriksen \& Widrow}{1999}]{b6} Henriksen R. N., Widrow L. M., 1999, MNRAS, 302, 321
\bibitem[\protect\citeauthoryear{Host \& Hansen}{2007}]{b30} Host O., Hansen S. H., 2007, JCAP, 06, 016
\bibitem[\protect\citeauthoryear{Host et al.}{2009}]{b18} Host O., Hansen S. H., Piffaretti R., Morandi A., Ettori S., Kay S.T., Valdarnini R., 2009, ApJ, 690, 358
\bibitem[\protect\citeauthoryear{Ishiyama}{2014}]{c39} Ishiyama, 2014, ApJ, 788, 271
\bibitem[\protect\citeauthoryear{Jing}{2000}]{b39} Jing Y. P., 2000, ApJ, 535, 30
\bibitem[\protect\citeauthoryear{Joyce, Marcos, \& Sylos Labini}{2009}]{2009MNRAS.397..775J} Joyce M., Marcos B., Sylos Labini F., 2009, MNRAS, 397, 775
\bibitem[\protect\citeauthoryear{Klypin et al.}{2014}]{c1} Klypin A., Yepes G., Gottlober S., Prada F., Hess S., 2014, preprint(arXiv:1411.4001)
\bibitem[\protect\citeauthoryear{Lacey \&  Cole}{1993}]{b8} Lacey C., Cole S., 1993, MNRAS, 262, 627
\bibitem[\protect\citeauthoryear{Ludlow et al.}{2012}]{b24} Ludlow A. D., Navarro J. F., Li M., Angulo R. E., Boylan-Kolchin M., Bett P. E., 2012, MNRAS, 427, 1322
\bibitem[\protect\citeauthoryear{Ludlow et al.}{2014}]{b13} Ludlow A. D., Navarro J. F., Angulo R. E., Boylan-Kolchin M., Springel V., Frenk Carlos., White S. D. M., 2014, MNRAS, 441, 378
\bibitem[\protect\citeauthoryear{Ludlow, Borzyszkowski \& Porciani}{2014}]{b40} Ludlow A. D., Borzyszkowski M., Porciani C., 2014, MNRAS, 445, 4110
\bibitem[\protect\citeauthoryear{Macci\`o, Dutton \& van den Bosch}{2008}]{b15} Macciò A. V., Dutton A. A., van den Bosch F. C., 2008, MNRAS, 391, 1940
\bibitem[\protect\citeauthoryear{Navarro, Frenk \& White}{1996}]{b2} Navarro J. F., Frenk C. S., White S. D. M.,1996, ApJ, 462, 563
\bibitem[\protect\citeauthoryear{Navarro, Frenk \& White}{1997}]{b3} Navarro J. F., Frenk C. S., White S. D. M.,1997, ApJ, 490, 493
\bibitem[\protect\citeauthoryear{Navarro et al.}{2004}]{c34} Navarro J. ~F., Hayashi E., Power C., Jenkins A. ~R., Frenk C. ~S., White S. ~D. ~M., Springel V., Stadel J., Quinn T. ~R., 2004, MNRAS, 349, 1039
\bibitem[\protect\citeauthoryear{Neto et al.}{2007}]{b38} Neto A. F. et al., 2007, MNRAS,381,1450
\bibitem[\protect\citeauthoryear{Padmanabhan}{1993}]{b27} Padmanabhan T., 1993, Structure Formation in the Universe
\bibitem[\protect\citeauthoryear{Peter}{2012}]{b5} Peter, A. H. G., 2012, preprint(arXiv:1201.3942) 
\bibitem[\protect\citeauthoryear{Prada et al.}{2012}]{b29} Prada F., Klypin A. A., Cuesta A. J., Betancort-Rijo J. E., Primack J., 2012, MNRAS, 423, 3018
\bibitem[\protect\citeauthoryear{Press \& Schechter}{1974}]{b1} Press W. H., Schechter, P., 1974, ApJ, 187, 425
\bibitem[\protect\citeauthoryear{Salvador-Sol{\'e} et al.}{2007}]{c12} Salvador-Sol{\'e} E., Manrique A., Gonz{\'a}lez-Casado G., Hansen S. H., 2007, ApJ, 666, 181
\bibitem[\protect\citeauthoryear{S{\'a}nchez-Conde \& Prada}{2014}]{b16} S{\'a}nchez-Conde M. A., Prada F., 2014, MNRAS, 442, 2271
\bibitem[\protect\citeauthoryear{Sheth, Mo \& Tormen}{2001}]{b35} Sheth R. K., Mo H. J., Tormen G., 2001, MNRAS, 321,1
\bibitem[\protect\citeauthoryear{Sparre \& Hansen}{2012}]{b19} Sparre M., Hansen S. H., 2012, JCAP, 10, 049
\bibitem[\protect\citeauthoryear{Voit}{2005}]{b32} Voit G. M., 2005, RvMP, 77,207
\bibitem[\protect\citeauthoryear{Zait, Hoffman \& Shlosman}{2008}]{b20} Zait A., Hoffman Y., Shlosman I., 2008, ApJ, 682, 835
\bibitem[\protect\citeauthoryear{Zavala \& Afshordi}{2015}]{2015arXiv150802713Z}  Zavala J., Afshordi N., 2015, preprint (arXiv:1508.02713)
\bibitem[\protect\citeauthoryear{Zeldovich}{1970}]{b36} Zeldovich Y. ~B. 1970, A\&A, 5, 84
\bibitem[\protect\citeauthoryear{Zhao et al.}{2003}]{b41} Zhao H.D., Jing Y. P., Mo H. J., B{\"o}rner G., 2003, ApJ, 597, L9 

\end{thebibliography}
\end{document}